\documentclass[
  aps,
  prx,
  twocolumn,
  superscriptaddress,
  longbibliography,
  nofootinbib
]{revtex4-2}

\usepackage{afterpage}

\usepackage[T1]{fontenc}
\usepackage[utf8]{inputenc}
\usepackage{amsmath,amssymb,amsfonts,bm}
\usepackage{mathtools}
\usepackage{physics}
\usepackage{graphicx}
\usepackage{algpseudocode}
\usepackage{tikz-cd}
\usepackage{ulem}
\usetikzlibrary{arrows.meta,positioning,fit,backgrounds}

\usepackage{orcidlink}

\newcommand{\E}{\mathbb{E}}
\newcommand{\cH}{\mathcal{H}}
\newcommand{\cD}{\mathcal{D}}
\newcommand{\cL}{\mathcal{L}}
\newcommand{\cM}{\mathcal{M}}

\newcounter{algorithm}

\usepackage{amsthm}
\newtheorem{theorem}{Theorem}

\newtheorem{proposition}{Proposition}
\newcommand{\R}{\mathcal R}
\newcommand{\id}{\mathrm{id}}

\begin{document}

\title{Reduced filtering using classical control hierarchy with linear optimality for continuous quantum error correction}

\author{Manuel Guatto \orcidlink{0009-0008-7043-1451}}

\affiliation{Forschungszentrum J\"ulich GmbH, Peter Gr\"unberg Institute, Quantum Control (PGI-8), 52425 J\"ulich, Germany}
\affiliation{Institute for Theoretical Physics, University of Cologne, D-50937 Cologne, Germany}

\author{F. A. C\'ardenas-L\'opez \orcidlink{0000-0002-2916-2826}}
\affiliation{Forschungszentrum J\"ulich GmbH, Peter Gr\"unberg Institute, Quantum Control (PGI-8), 52425 J\"ulich, Germany}

\author{Felix Motzoi \orcidlink{0000-0003-4756-5976}}
\affiliation{Forschungszentrum J\"ulich GmbH, Peter Gr\"unberg Institute, Quantum Control (PGI-8), 52425 J\"ulich, Germany}
\affiliation{Institute for Theoretical Physics, University of Cologne, D-50937 Cologne, Germany}

\date{\today}

\begin{abstract}
Continuous quantum error correction promises reduced thresholds and hardware overhead compared to the digital paradigm, but its complex model estimation has greatly reduced its practicality. We recast this problem as a classical control-theoretic state hierarchy made of an exact bilinear reduced model, an exact conditional filter, and standard estimation and feedback design. For any $[[n,k,d]]$ stabilizer code whose generators are all continuously monitored and which is driven by a single Pauli recovery channel, we prove that the exact filter closes at dimension $3\cdot2^{\,n-k-1}$. This dimension is set by the syndrome space instead of the Hilbert space and is minimal within invariant-subspace filtering. The reduced model exposes and corrects two failure mechanisms of the feedback loop. Strong monitoring damps the rotational partner through which the controller steers the stabilizer and injects back-action diffusion along the recovery path; lowering the measurement strength during correction raises the entanglement fidelity of partner-steering LQR control by $0.096$ in the three-qubit code with three monitored parities. Linearizing at the codespace leaves the syndrome uncontrollable, so the controller can stall; a gain designed at a shifted operating point removes this degeneracy. In the limit of zero measurement during correction, we prove that a syndrome-only filter is exact at every recovery boundary. The resulting committed-recovery controller needs only $2^{\,n-k}$ class probabilities and gives the highest entanglement fidelity compared among the methods tested in the literature, $0.947$ against $0.920$ for a measurement-based estimator protocol in the three-qubit bit-flip code, and $0.726$ vs.~at most $0.407$ for continuous controllers in the five-qubit perfect code.
\end{abstract}

\maketitle

\section{Introduction}

Continuous quantum error correction (CQEC) replaces discrete syndrome-extraction rounds with weak, continuous monitoring of the stabilizers and corrective Hamiltonian feedback~\cite{ahn_continuous_2002,ahn_quantum_2003,sarovar_practical_2004,sarovar_cooling_2005}. Superconducting circuits support continuous weak measurement with real-time feedback~\cite{vijay_stabilizing_2012} and low-latency feedback electronics~\cite{salathe_lowlatency_2018,Reuer_2023}, and a three-qubit bit-flip code has been operated with continuous parity measurements~\cite{livingston_experimental_2022}. The syndrome currents are noisy, so the controller must filter them, and the conditional state that an exact filter propagates grows exponentially with the number of qubits~\cite{wiseman_milburn_2009,jacobs_steck_2006,bouten_introduction_2007}. The measurement strength adds a further difficulty: stronger monitoring sharpens detection, but it also disturbs the coherent recovery it is meant to trigger.

\textit{Relation to prior CQEC work.}
Ahn, Doherty, and Landahl formulated CQEC for the three-qubit bit-flip code with continuous measurement, stochastic state estimation, and bang-bang Hamiltonian feedback that maximizes the rate of return to the codespace~\cite{ahn_continuous_2002}. Their simulations integrate the full conditioned density matrix, and they showed that, for a maximally mixed codespace prior, $(2^{\,n-k})^2$ parameters (the populations of the error spaces and the coherences between them) suffice for arbitrary controls; feedback constructions for continuously detected errors were extended to several error channels per qubit~\cite{ahn_quantum_2003,ahn_quantum_2004}. Van Handel and Mabuchi showed that syndrome probabilities can be tracked exactly by a low-dimensional classical filter in the absence of feedback~\cite{vanhandel2005}, and Chase, Landahl, and Geremia built approximate reduced filters that include recovery~\cite{chase_efficient_2008}. Mabuchi interpreted CQEC as a classical hybrid control problem~\cite{mabuchi_continuous_2009}. Later work studied Bayesian and boxcar filters for continuous parity records~\cite{mohseninia_alwayson_2020,convy_logarithmic_2022}, double-threshold protocols~\cite{atalaya_timedependent_2021}, noncommuting measurements in the Bacon-Shor code~\cite{atalaya_baconshor_2020}, machine-learned detectors~\cite{convy_machine_2022}, quantum-jump CQEC with weak corrections~\cite{hsu_method_2016}, noise-assisted feedback~\cite{cardona_continuous_2019}, perturbative analyses~\cite{ippoliti_perturbative_2015}, and learned codes with weak recoveries~\cite{lanka_optimizing_2025}. Borah \textit{et al.} proposed a measurement-based estimator that propagates an auxiliary quantum state driven by the stabilizer currents and corrects from that online estimate~\cite{borah_measurement_2022}. Autonomous schemes avoid the classical loop altogether through coherent feedback~\cite{kerckhoff_designing_2010}. On the hardware side, Halaski and Koch showed that parity meters built from two-body couplings dephase the logical information, and noted that pausing the measurement drive before a correction can reduce this damage~\cite{halaski_obstacles_2026}. Control-theoretic studies have also addressed stabilizability of open quantum dynamics~\cite{linowski_spectral_2023} and the exponential stability of invariant subspaces under quantum stochastic master equations~\cite{benoist_exponential_2017}.

We organize the problem as a hierarchy of control-oriented reduced models. Exact expectation-value dynamics on a compact operator basis are specialized to syndrome sectors, where they become bilinear control systems~\cite{elliott_bilinear_2009,altafini_modeling_2012}, and the feedback is synthesized with standard tools from classical control theory: local LQR design~\cite{anderson_optimal_1990}, regularized dynamic inversion~\cite{enns_dynamic_1994}, and interacting-multiple-model (IMM) belief updates~\cite{blom_interacting_1988}. The estimator, the sector selector, and the controller can be exchanged independently inside one pipeline, and the online laws use relatively few expectation values and small matrices.

The reduction also serves as a diagnostic. Because the conditional dynamics are exact on a few interpretable coordinates (observables), the loop can be linearized, its controllability computed, and fast variables eliminated adiabatically. We use this to separate two mechanisms behind the difficulty of choosing a measurement strength, and to remove each with a classical intervention that needs no change to the hardware. Our main results are the following.

(i) Theorem~\ref{thm:closure}: for any stabilizer code with continuously monitored generators and a single Pauli recovery channel, the exact filter carries $3\cdot2^{\,n-k-1}$ coordinates, the stabilizer group together with half of its coset under the control. This realizes the minimal invariant subspace of Ref.~\cite{GrigolettoPellegriniTicozzi2025} in closed form.
(ii) Measurement back-action along the recovery path, which grows as $\sqrt{\kappa}$, competes with syndrome information, which accumulates at rate $\eta\kappa$; scheduling $\kappa$ between detection and correction resolves the trade for partner-steering controllers (Sec.~\ref{sec:tradeoff}).
(iii) Codespace-linearized LQR (Linear Quadratic Regulator) has a controllability degeneracy that can stall the estimator--controller cascade; a shifted operating point removes it (Sec.~\ref{sec:deadlock}).
(iv) Theorem~\ref{thm:shutdown}: with the measurement switched off during a committed Pauli rotation and covariant noise, a syndrome-only filter is exact at every recovery boundary (Sec.~\ref{sec:shutdown}).
(v) Benchmarks on the three-qubit and five-qubit codes against implementations of the Ahn, Chase--Landahl--Geremia, and Borah approaches (Secs.~\ref{sec:three-unified} and~\ref{sec:fivequbit}).

The closure construction, the back-action versus information trade, and the prediction that measurement scheduling helps filter-based controllers while penalizing open-loop recovery follow from Eqs.~\eqref{eq:sme} and~\eqref{eq:measurement-currents} and hold for both codes studied here. The deadlock of Sec.~\ref{sec:deadlock} is derived in general but is observable only at strong measurement with errors sparse enough that the register idles between corrections. We use the three-qubit code as the working example because it is the smallest code in which a syndrome sector is a two-dimensional rotation plane, so the mechanisms can be written in closed form. However the framework can be applied to any QEC code.

\section{Control-oriented model hierarchy}
\label{sec:methods}

\subsection{Feedback-loop architecture}

\begin{figure*}[t]
\centering
\includegraphics[width = \linewidth]{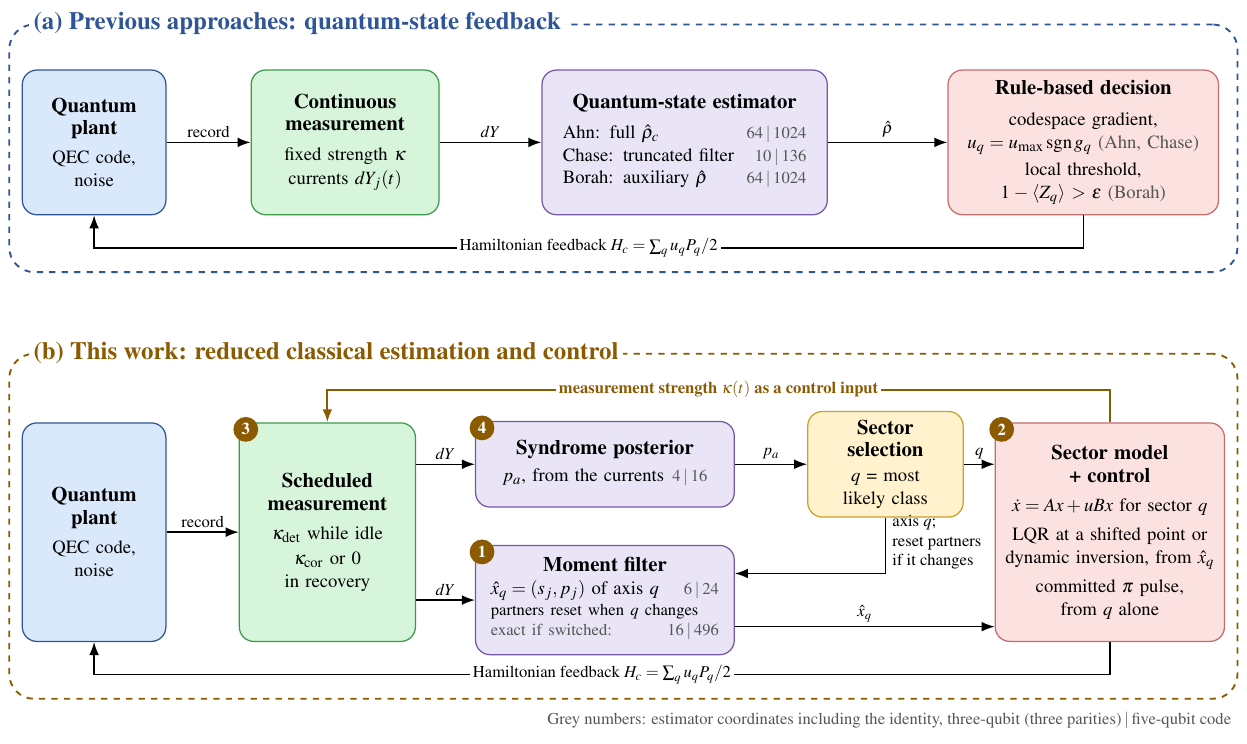}
\caption{Comparison with earlier continuous-QEC feedback loops. (a) Earlier approaches, as implemented in Secs.~\ref{sec:three-unified} and~\ref{sec:fivequbit}, propagate a full or truncated quantum-state estimate at fixed measurement strength and act on it with a decision rule: the codespace-population gradient of Refs.~\cite{ahn_continuous_2002,chase_efficient_2008} or the local-$Z$ threshold of Ref.~\cite{borah_measurement_2022}. (b) This work. A syndrome posterior selects the error sector $q$, which sets both the controller and the active axis of the moment filter. Steps: (1) For a fixed axis the moment filter is exact on the $3\cdot2^{\,n-k-1}$ coordinates of Theorem~\ref{thm:closure}; when $q$ changes, its partner moments are reset, whereas an exact switched filter would need the full common closure. (2) Feedback is designed on exact sector-wise bilinear models (Sec.~\ref{subsec:lqr}). (3) The measurement strength is a control input: it is lowered to $\kappa_{\rm cor}$ while a continuous controller corrects, and switched off during a committed pulse (Secs.~\ref{sec:tradeoff} and~\ref{sec:shutdown}). (4) The calibrated syndrome posterior of the shutdown controller is exact while no drive acts and, by Theorem~\ref{thm:shutdown}, at the end of every committed pulse when the measurement is fully off and the noise is covariant; since the posterior receives no record during the pulse, the axis $q$ stays fixed until the rotation is complete. Grey numbers give number of estimator coordinates, including the identity, for the three-qubit code with three monitored parities and the five-qubit code (Secs.~\ref{sec:three-unified}, \ref{sec:fivequbit} and~\ref{sec:native-robustness}).}
\label{fig:feedback-loop}
\end{figure*}

Fig.~\ref{fig:feedback-loop} shows the closed feedback loop. The encoded register (the plant) evolves under physical noise, diffusive stabilizer measurements, and local Hamiltonian control. The measured outputs are noisy syndrome currents, which do not reveal the syndrome at any single instant and must be filtered. Two estimators run on the same currents: a posterior over the syndrome classes (Sec.~\ref{subsec:imm_approach}) and a reduced moment filter (Sec.~\ref{subsec:conditional-filter}). At each decision interval the posterior selects the most likely error sector. That choice fixes the moment filter's active recovery axis and the sector-wise bilinear model (Sec.~\ref{subsec:bilinear}), from which the controller computes a bounded Hamiltonian command, either by a local LQR design or by regularized dynamic inversion, and applies it back to the plant. Two elements of the loop are left as scheduled degrees of freedom and become central in Secs.~\ref{sec:tradeoff} and~\ref{sec:deadlock}: the measurement strength $\kappa(t)$, which may be switched by the posterior sector estimate, and the linearization point of the LQR design.

\subsection{Physical model and stochastic master equation}

We present the basic model for the three-qubit bit-flip code on $\cH=(\mathbb{C}^2)^{\otimes 3}$, whose stabilizer group $\mathcal{S}$ is generated by
\begin{equation}
S_{12}=Z_1Z_2, \qquad S_{23}=Z_2Z_3.
\end{equation}
Both generators are monitored diffusively with measurement strength $\kappa>0$ and detector efficiency $\eta\in(0,1]$. Bit-flip errors act on each qubit at rate $\gamma\ge 0$, and corrections are applied through
\begin{equation}
H_c(t)=\frac{1}{2}\sum_{i=1}^{3} u_i(t) X_i,
\label{eq:control-hamiltonian}
\end{equation}
one channel per single-qubit recovery, the factor $1/2$ making
$u_i$ the angular rate of the syndrome rotation it generates
(Sec.~\ref{subsec:bilinear}). For a general code the recoveries are
the Pauli operators prescribed by the decoder, and
Eq.~\eqref{eq:control-hamiltonian} is replaced by
$H_c(t)=\tfrac12\sum_{a}u_a(t)P_a$ over the available correction
channels $P_a$; the sector-wise construction of
Sec.~\ref{subsec:bilinear} and Theorem~\ref{thm:closure} use the
single active channel $H_q=\tfrac12 u P$, and the five-qubit
instance is Eq.~\eqref{eq:5q-control}. Conditioned on the measurement record, the state obeys the diffusive stochastic master equation~\cite{wiseman_milburn_2009}
\begin{equation}
\begin{split}
d\rho={}\cL_u(\rho)dt +\sum_{j\in\{12,23\}}\sqrt{\eta\kappa}\,\cM[S_j]\rho\, dW_j,
\end{split}
\label{eq:sme}
\end{equation}
with independent Wiener increments $dW_j$, the measurement superoperator
\begin{equation}
\cM[c]\rho=c\rho+\rho c^{\dagger}-\Tr[(c+c^{\dagger})\rho]\rho,
\end{equation}
and the Lindbladian
\begin{equation}
\cL_u(\rho)= -i[H_c(u),\rho]
+ \sum_{i=1}^{3}\gamma\,\cD[X_i]\rho
+ \sum_{j\in\{12,23\}}\kappa\,\cD[S_j]\rho,
\label{eq:lindbladian}
\end{equation}
where $\cD[L]\rho=L\rho L^{\dagger}-\tfrac{1}{2}\qty(L^{\dagger}L\rho+\rho L^{\dagger}L)$. The first term of Eq.~\eqref{eq:lindbladian} is the control, the second the physical noise, and the third the measurement-induced dephasing that accompanies Eq.~\eqref{eq:sme}. The control amplitudes depend on filtered state estimates rather than on the instantaneous measurement currents\cite{chen2020PI,martin2015hybrid} . Consequently, the feedback enters through the Hamiltonian increment $H_c(u(t))\,dt$, and the additional Wiseman–Milburn terms associated with direct current feedback \cite{wiseman_milburn_2009} do not arise. The measurement currents are
\begin{equation}
dY_j(t)=2\sqrt{\eta\kappa}\,\ev{S_j}_{\rho}dt+dW_j(t),
\qquad j\in\{12,23\},
\label{eq:measurement-currents}
\end{equation}
with $\ev{O}_\rho=\Tr(O\rho)$. Equal error rates on the three qubits are a benchmark choice. The estimator uses its own innovation $dY_j-2\sqrt{\eta\kappa}\,\ev{S_j}_{\hat\rho}dt$ and never receives hidden error labels or the reference state.

\subsection{Exact reduced model for a stabilizer code}
\label{subsec:bilinear}

For a code with monitored generators $\{g_1,\dots,g_{n-k}\}$, Pauli noise operators $L_\mu$ at rates $\gamma_\mu$, and control Hamiltonian $H_c(u)$, the unconditional expectation of any observable $O$ obeys
\begin{equation}
\begin{split}
\frac{d}{dt}\ev{O}={}&\ev{\cL_u^{\dagger}(O)} \\
={}&i\ev{[H_c(u),O]}
+\sum_{\mu}\gamma_\mu\ev{\cD^{\dagger}[L_\mu]O} \\
&+\sum_{j}\kappa_j\ev{\cD^{\dagger}[g_j]O},
\end{split}
\label{eq:heis-general}
\end{equation}
with $\cD^{\dagger}[L]O=L^{\dagger}OL-\frac{1}{2}\{L^{\dagger}L,O\}$. The commutator is the only source of control; the dissipators give uncontrolled damping. We write $\cdot|_H$ for the commutator term alone.

Take a syndrome class with representative error $E$, and let
\begin{equation}
  F = \{\, j : \{E, g_j\} = 0 \,\}, \qquad w := |F| \le n-k,
  \label{eq:flipped-set}
\end{equation}
be the set of monitored generators it flips. The recovery $P$ assigned by the decoder has the same syndrome, so $\{P,g_j \} = 0$ exactly for $j \in F$, and it is applied through $H_q = \frac{1}{2} u P$ with a single scalar amplitude. For $j\in F$ the commutator $i[H_q,g_j]$ is proportional to $iPg_j\notin\mathcal{S}$, so the description cannot close on the stabilizer coordinates $s_j := \ev{g_j}$ alone. We therefore define the \textit{rotational partners}
\begin{equation}
  \tilde g_j := i\, P g_j , \qquad j \in F .
  \label{eq:partner-def}
\end{equation}
Anticommutation gives $\tilde g_j^{\dagger} = \tilde g_j$ and $\tilde g_j^{\,2} = I$, so $p_j := \ev{\tilde g_j} \in [-1,1]$ is a coordinate on the same footing as $s_j$. The same relation gives
\begin{align}
  &i[H_q, g_j] = u_j\, \tilde g_j ,
  \label{eq:plane-obstruction}
  \\
  &i[H_q, \tilde g_j] = -u_j\, g_j ,
  \label{eq:plane-commutators}
\end{align}
the second using $P g_j P = -g_j$. Eq.~\eqref{eq:plane-obstruction} takes the description out of $\mathcal{S}$, and Eq.~\eqref{eq:plane-commutators} returns $g_j$ instead of a third observable, so adjoining the partner is enough. Each flipped generator contributes one exact two-dimensional rotation plane, and a sector is the direct sum of $w$ such planes driven by the same $u$. The partners are exactly the coset elements $P\cdot\mathcal{S}^-$ that Theorem~\ref{thm:closure} below adjoins to $\mathcal{S}$.

Collecting $x = (\dots, s_j, p_j, \dots)^{\top}$ and taking expectations in Eq.~\eqref{eq:heis-general} gives the exact bilinear system
\begin{equation}
  \dot{x} = A x + \sum_{j} u_j(t)\, B_j x .
  \label{eq:bilinear-full}
\end{equation}
Equation~\eqref{eq:bilinear-full} is exact. It is not a cumulant truncation or a small-signal expansion: the closure is a property of the code's Pauli algebra, and the higher moments that such approximations discard never arise. The system is linear in $x$ at fixed $u$ and linear in $u$ at fixed $x$, but not jointly linear in $(x,u)$.

Its two matrices are the two Jacobians of the vector field $f(x,u) = Ax + \sum_j u_j B_j x$. The state Jacobian $\partial f/\partial x = A + \sum_j u_j B_j$ reduces to $A$ at zero control and, because $f$ is linear in $x$ at fixed $u$, does not depend on the point where it is evaluated: $A$ is an exact  Jacobian. The input Jacobian $\partial f/\partial u_j = B_j x$ depends on the state; $B_j$ itself is the constant mixed derivative $\partial^2 f/\partial x\,\partial u_j$, the coefficient matrix of the linear control vector field $g_j(x) = B_j x$. 

Two properties follow and are used later. First, the plant is unconditional: it is the average over measurement records, so the efficiency $\eta$ does not enter it and $\kappa$ appears only as damping. Gain synthesis needs nothing more, since the Riccati equation~\eqref{eq:Ricatti} uses only $(A_i,b_{i,\mathrm{lin}})$. Second, the loop is designed on Eq.~\eqref{eq:bilinear-full} and run on the conditional filter of Sec.~\ref{subsec:conditional-filter}. That separation would be exact for a linear plant with additive noise. Here it is not, because the input Jacobian depends on the state and the filter innovation is multiplicative in the retained coordinates; Secs.~\ref{sec:tradeoff} and~\ref{sec:deadlock} analyze the consequences.

Both matrices are read off the Pauli algebra. The control matrix of a recovery $P$ follows from Eqs.~\eqref{eq:plane-obstruction} and~\eqref{eq:plane-commutators}: in each plane $(s_j,p_j)$ with $j\in F$ it has the entries $+1$ in row $s_j$, column $p_j$ and $-1$ in row $p_j$, column $s_j$, and it is zero elsewhere. The drift matrix is diagonal, $A = -\operatorname{diag}(a_{O_1},\dots,a_{O_N})$ because for a Pauli $L$ with $L^2 = I$,
\begin{equation}
  \cD^{\dagger}[L] O = L O L - O =
  \begin{cases}
    0    & [L,O] = 0, \\
    -2 O & \{L,O\} = 0,
  \end{cases}
  \label{eq:pauli-dissipator}
\end{equation}
so every Pauli observable is an eigenvector of the drift, with decay rate:
\begin{equation}
  a_O = a_O^{\rm noise} + 2\kappa\, m_O ,
  \label{eq:drift-rule}
\end{equation}
\begin{equation}
  a_O^{\rm noise} := 2 \!\!\sum_{\mu : \{L_\mu, O\} = 0}\!\! \gamma_\mu ,
  \qquad
  m_O := \bigl| \{l : \{g_l, O\} = 0\} \bigr| ,
  \label{eq:drift-split}
\end{equation}
one contribution each for the anticommuting noise channels and the anticommuting monitored generators, [Eqs.~\eqref{eq:heis-general} and~\eqref{eq:pauli-dissipator}]. Because $\mathcal S$ is abelian, $m_{g_j}=0$: the stabilizer coordinates undergo no measurement damping. For the partners, pushing a generator through Eq.~\eqref{eq:partner-def} gives
\begin{equation}
  g_l (P g_j) = (g_l P) g_j = \pm (P g_l) g_j = \pm (P g_j) g_l ,
\end{equation}
where the last step again uses that $\mathcal{S}$ is abelian and the sign is that of $g_l P$ against $P g_l$. Hence $\tilde g_j$ anticommutes with $g_l$ if and only if $P$ does, that is, if and only if $l\in F$, independently of $j$. Every partner in the sector therefore sees the same $w$ anticommuting generators, $m_{\tilde g_j} = w$, and all partners carry identical measurement damping. Collecting the two rules, the sector model is, for each $j \in F$,
\begin{equation}
  \begin{split}
    \dot s_j &= -a_s^{\rm noise}\, s_j + u\, p_j , \\
    \dot p_j &= -\bigl(a_p^{\rm noise} + 2\kappa w\bigr) p_j - u\, s_j .
  \end{split}
  \label{eq:sector-general}
\end{equation}
The damping of the control lever is set by the sector multiplicity $w$, which returns in Sec.~\ref{sec:deadlock}. The noise rates $a_s^{\rm noise}$ and $a_p^{\rm noise}$ differ only through channels acting on the corrected qubit; they coincide for depolarizing noise and for the bit-flip code. The control map is skew-symmetric, $x^\top B x = 0$, so each actuator rotates its planes at angular rate $u$ while the drift damps the partners faster than the stabilizers. In a filter that names its observables once for all recoveries, a partner can differ from Eq.~\eqref{eq:partner-def} by a sign $c_j=\pm1$; Appendix~\ref{app:polar} provides the rule. The sign does not change the plant but enters any feedback law that sums partner estimates.

\subsection{The three-qubit case}
\label{sec:sector}
For the bit-flip code, $g_1 = Z_1Z_2$, $g_2 = Z_2Z_3$, the recoveries are $X_i$, and the noise is $X$-type at rate $\gamma$. The commutator $i[X_i,g_j]$ equals twice $g_j$ with $Z_i$ replaced by $Y_i$ when qubit $i$ lies in the support of $g_j$, and vanishes otherwise. The four partners are therefore
\begin{equation}
  \tilde g \in \{\, Y_1Z_2,\; Z_1Y_2,\; Y_2Z_3,\; Z_2Y_3 \,\},
  \label{eq:3q-partners}
\end{equation}
and the stabilizer rates read
\begin{align}
  \dot{s}_{12}\big|_{H} &= u_1\ev{Y_1Z_2} + u_2\ev{Z_1Y_2}, \\
  \dot{s}_{23}\big|_{H} &= u_2\ev{Y_2Z_3} + u_3\ev{Z_2Y_3}.
  \label{eq:stabilizer-control-open}
\end{align}
Each basis operator anticommutes with two of the three errors, so $a^{\rm noise} = 4\gamma$ throughout. For a candidate error sector $i \in \{1,2,3\}$,
\begin{equation}
  \dot x^{(i)} = A_i x^{(i)} + u_i B_i x^{(i)} .
  \label{eq:sector-bilinear}
\end{equation}
Sectors 1 and 3 have $w = 1$:
\begin{align}
  x^{(1)} &= \begin{pmatrix} s_{12} \\ p_1 \end{pmatrix}
           = \begin{pmatrix} \ev{Z_1Z_2} \\ \ev{Y_1Z_2} \end{pmatrix},
  \\[2pt]
  x^{(3)} &= \begin{pmatrix} s_{23} \\ p_4 \end{pmatrix}
           = \begin{pmatrix} \ev{Z_2Z_3} \\ \ev{Z_2Y_3} \end{pmatrix},
  \\[2pt]
  A_1 &= A_3 = \begin{pmatrix} -4\gamma & 0 \\ 0 & -(4\gamma+2\kappa)
              \end{pmatrix},
  \\[2pt]
  B_1 &= B_3 = \begin{pmatrix} 0 & 1 \\ -1 & 0 \end{pmatrix}.
\end{align}
Sector 2 has $w = 2$, because qubit 2 lies in the support of both generators:
\begin{align}
  x^{(2)} &= \bigl( s_{12},\, p_2,\, s_{23},\, p_3 \bigr)^{\top}
  \notag \\
          &= \bigl( \ev{Z_1Z_2},\, \ev{Z_1Y_2},\, \ev{Z_2Z_3},\,
                    \ev{Y_2Z_3} \bigr)^{\top},
  \\[2pt]
  A_2 &= \operatorname{diag}\bigl( -4\gamma,\, -(4\gamma+4\kappa),
  \notag \\
      &\hspace{4.2em} -4\gamma,\, -(4\gamma+4\kappa) \bigr),
  \\[2pt]
  B_2 &= \begin{pmatrix}
           0 & 1 & 0 & 0 \\ -1 & 0 & 0 & 0 \\
           0 & 0 & 0 & 1 \\ 0 & 0 & -1 & 0
         \end{pmatrix}.
\end{align}
All three sectors have $c_j = +1$, because the generators are $Z$-type and the recoveries $X$-type. The five-qubit code of Sec.~\ref{sec:fivequbit} realizes multiplicities up to $w = 4$ and both signs.

\subsection{Exact conditional filter and its minimal dimension}
\label{subsec:conditional-filter}
The loop runs on an estimator driven by the measurement record. We now show that this estimator is finite dimensional for every stabilizer code under a single Pauli recovery, with a dimension fixed by the syndrome register.

Choose an operator family $\mathcal{B}=\{O_a\}_{a=1}^{N}$ containing the monitored generators and define $x_a(t):=\ev{O_a}_{\rho_c}$. It\^o calculus gives
\begin{equation}
\begin{split}
dx_a ={}& \ev{\mathcal{L}^{\dagger}(O_a)}dt \\
&+\sum_{j=1}^{m}\sqrt{\eta_j\kappa_j}\,
\Bigl(
\ev{S_jO_a+O_aS_j}
-2\hat{s}_j x_a
\Bigr)dW_j.
\end{split}
\label{eq:general-coordinate-filter}
\end{equation}
The drift is the object of Sec.~\ref{subsec:bilinear}. A finite filter exists when both the drift and the innovation coefficient $\ev{S_jO_a+O_aS_j}$ are functions of the retained coordinates. Closure thus involves three operations: control multiplies by the recovery when the two anticommute [Eqs.~\eqref{eq:plane-obstruction} and~\eqref{eq:plane-commutators}], Pauli drift returns $0$ or $-2O$ and never enlarges the basis [Eq.~\eqref{eq:pauli-dissipator}], and the measurement update maps $O\mapsto S_jO+OS_j$. Seeding with $\mathcal{B}_0=\{S_1,\dots,S_m\}$, one iterates
\begin{equation}
\begin{split}
\mathcal{B}_{k+1} ={}& \mathcal{B}_k
\cup \qty{i[H_{\ell},O]: O\in\mathcal{B}_k} \\
&\cup \qty{\mathcal{D}^{\dagger}[L_{\mu}]O: O\in\mathcal{B}_k} \\
&\cup \qty{S_jO+OS_j: O\in\mathcal{B}_k},
\end{split}
\label{eq:general-basis-recursion}
\end{equation}
over controls $\ell$, noise operators $\mu$, and measurement channels $j$, and stops at the first $K$ with $\operatorname{span} (\mathcal{B}_{K+1})=\operatorname{span}(\mathcal{B}_{K})$. The recursion always terminates, since the operator space has dimension at most $4^n-1$, but in general its fixed point must be computed numerically. For a single Pauli control it has a closed form.

We work with Pauli operators modulo phase, so any two observables commute or anticommute. Let $\mathcal{S}=\langle g_1,\dots,g_{n-k}\rangle$, with $|\mathcal{S}|=2^{n-k}$, and write $\mathcal{S}^-=\{s\in\mathcal{S}:\{P,s\}=0\}$ and $\mathcal{S}^+=\mathcal{S}\setminus\mathcal{S}^-$.

\begin{theorem}[Closure dimension]
\label{thm:closure}
Let all $n-k$ stabilizer generators be continuously monitored, let the noise operators $L_\mu$ be Pauli operators, and let the control set consist of a single Pauli $P$ that anticommutes with at least one element of $\mathcal{S}$. Then the recursion~\eqref{eq:general-basis-recursion} seeded by $\mathcal{B}_0=\{g_1,\dots,g_{n-k}\}$ terminates at
\begin{equation}
\mathcal{B}=\bigl(\mathcal{S}\setminus\{I\} \bigr)\;\cup\;P\cdot\mathcal{S}^-,
\label{eq:closure-set}
\end{equation}
of cardinality
\begin{equation}
|\mathcal{B}|
=\bigl(2^{\,n-k}-1\bigr)+2^{\,n-k-1}
=3\cdot2^{\,n-k-1}-1 .
\label{eq:closure-dimension}
\end{equation}
Including the identity, the closed span has dimension $3\cdot2^{\,n-k-1}=\tfrac{3}{2}|\mathcal{S}|$.
\end{theorem}

The proof is in Appendix~\ref{app:closure-proof}. Noise never enlarges the basis, and the measurement update and the control act as multiplication by a generator and by $P$, each allowed only under a commutation condition; the fixed point is the orbit of the seeds under this partial action. Throughout the paper we quote the identity-included dimension $3\cdot2^{\,n-k-1}$ of the propagated linear filter.

Three consequences follow from the theorem. First, the dimension depends only on the number $n-k$ of monitored generators. We verified numerically that $[[3,1,1]]$, $[[4,2,2]]$ and $[[6,4,2]]$ all return dimension $6$. Second, since $2^{\,n-k}$ is the number of syndromes, the exact filter carries three coordinates for every two syndromes. Its state is of the same order as the syndrome register any decoder maintains, and exponentially smaller than the $2^{2n}-1$ parameters of the conditional density operator (Table~\ref{tab:closure-dimensions}). Third, the hypothesis on $P$ holds for any single-qubit control on a code with $d\ge2$, because a $P$ commuting with all of $\mathcal{S}$ would lie in the normalizer $N(\mathcal{S})$: it cannot lie in $N(\mathcal{S})\setminus\mathcal{S}$, whose elements have weight at least $d$, and a weight-one stabilizer would freeze one qubit, which none of the codes considered here does. The bit-flip code, with $d=1$, satisfies it only for the $X$-type controls used here.

\paragraph*{Minimality.}
The recursion~\eqref{eq:general-basis-recursion} is an instance of the invariant-subspace reduction of Grigoletto, Pellegrini, and Ticozzi~\cite{GrigolettoPellegriniTicozzi2025}\cite{Grigoletto2025exactmodelreduction}: it generates the minimal real operator span that contains the monitored generators and the identity and is invariant under the adjoint generator and the measurement maps. Theorem~\ref{thm:closure} gives that span in closed form, with its coset structure, for a whole family of codes; Ref.~\cite{GrigolettoPellegriniTicozzi2025} establishes the reduction and computes it numerically. The minimality holds within this construction. It does not cover nonlinear estimators, restricted initial states, or the count of independent coordinates once normalization is used.

\paragraph*{Scope of the closed form.}
\label{subsec:closure-scope}
Theorem~\ref{thm:closure} concerns one active control channel, which matches our architecture: the recovery law controls one Pauli channel at a time, selected by the syndrome estimate. 

Switching dynamically for any of these syndrome-specific $P$ to $P'$ generally needs moments outside this new branch to initialize it exactly. Closing the recursion under all permitted controls in $P$ and $P'$ gives 14 nonidentity moments for the three-qubit code with two monitored generators (15 when $Z_1Z_3$ is also monitored) and 495 for the five-qubit code, close to half of the 1023 parameters of its density operator. The closure makes no assumption on the encoded state; with the maximally mixed codespace prior used by our estimators, the $(2^{\,n-k})^2$ syndrome-space parameters of Ref.~\cite{ahn_continuous_2002} (16 and 256) already suffice for arbitrary controls. Our switched estimators therefore apply a declared sector projection when the active axis changes: the stabilizer-group moments are kept and the new partners are set to zero, which is the restriction of
\begin{equation}
 \widehat\rho\longmapsto\sum_a\Pi_a\widehat\rho\Pi_a.
 \label{eq:sector-projection}
\end{equation}
This removes inter-sector coherence without changing syndrome probabilities when the control operator changes; it is an approximation. The benchmarks of Secs.~\ref{sec:three-unified} and~\ref{sec:fivequbit} propagate exactly the fixed-channel span of Theorem~\ref{thm:closure}, six coordinates for the three-qubit code and twenty-four for the five-qubit code, with this projection at axis changes. A nine-coordinate model that carries the partners of all three recoveries at once was used in an earlier timing study and is given in Appendix~\ref{app:9dfilter}. Theorem~\ref{thm:shutdown} of Sec.~\ref{sec:shutdown} identifies an architecture in which a syndrome-only filter is exact at every recovery boundary. Finally, the scaling $3\cdot2^{\,n-k-1}$ is exponential in the number of monitored generators. It matches the syndrome space, the natural reference for syndrome-based schemes, but not the polynomial cost of matching decoders for topological codes.

\begin{table}[t]
\caption{Exact filter dimension of Theorem~\ref{thm:closure} against the conditional density operator, for the codes on which Eq.~\eqref{eq:closure-set} has been verified as a set identity. The dimension tracks the syndrome count $2^{\,n-k}$ and is independent of $n$.}
\label{tab:closure-dimensions}
\begin{ruledtabular}
\begin{tabular}{lccccc}
Code & $n$ & $n-k$ & $3\cdot2^{n-k-1}$ & syndromes & $2^{2n}-1$ \\
\hline
$[[3,1,1]]$ bit-flip & 3 & 2 & 6   & 4   & 63 \\
$[[4,2,2]]$          & 4 & 2 & 6   & 4   & 255 \\
$[[6,4,2]]$          & 6 & 2 & 6   & 4   & 4095 \\
$[[5,1,3]]$ perfect  & 5 & 4 & 24  & 16  & 1023 \\
$[[8,3,3]]$          & 8 & 5 & 48  & 32  & 65535 \\
$[[7,1,3]]$ Steane   & 7 & 6 & 96  & 64  & 16383 \\
$[[9,1,3]]$ Shor     & 9 & 8 & 384 & 256 & 262143 \\
\end{tabular}
\end{ruledtabular}
\end{table}

\subsection{Sector estimation}
\label{subsec:imm_approach}
The filter estimates the monitored stabilizers continuously, but the recovery law must decide which Pauli channel to drive. The sector estimator keeps a belief $p(t)=\bigl(p_\sigma(t)\bigr)_\sigma$ over the $2^{n-k}$ syndrome classes and passes the feedback law either the maximum-a-posteriori class or the full belief. Each class $\sigma$ has a template $\mu_\sigma\in\{\pm1\}^{\,n-k}$, the stabilizer values of the codeword hit by its representative error. The belief is updated in two steps, as in an IMM estimator~\cite{blom_interacting_1988}.

\paragraph*{Prior update.} Let $\{E_a\}$ be the correctable errors, with rates $\gamma_a$ and syndromes $s(E_a)\in\{\pm1\}^{\,n-k}$, so that $E_a$ carries class $\sigma$ to $\sigma\cdot s(E_a)$ componentwise. Between updates the belief follows the continuous-time Markov chain
\begin{equation}
\dot p_{\sigma}
= -\Bigl(\sum_a \gamma_a\Bigr) p_{\sigma}
+ \sum_a \gamma_a\, p_{\sigma\cdot s(E_a)} ,
\label{eq:hazard}
\end{equation}
a symmetric random walk on the syndrome graph with total escape rate $\sum_a\gamma_a$.

\paragraph*{Likelihood update.} Over a step $dt$ the currents $dY=(dY_1,\dots,dY_{n-k})$ are conditionally Gaussian [Eq.~\eqref{eq:measurement-currents}], with a mean set by the stabilizer values of the class. Each class receives the tempered likelihood
\begin{equation}
\mathcal{L}_\sigma(dY)
\propto
\exp\Bigl(-\frac{\beta}{2\,dt}\,
\norm{dY-2\sqrt{\eta\kappa}\,c\,\mu_\sigma\,dt}_2^2\Bigr),
\label{eq:imm-likelihood}
\end{equation}
followed by $p_\sigma(t^+)\propto p_\sigma(t^-)\,\mathcal{L}_\sigma(dY)$ and normalization. The contrast $c\in(0,1]$ accounts for stabilizers that are not exactly $\pm1$ during transients, and the tempering exponent $\beta\in(0,1]$ prevents a single noisy increment from forcing a switch. We use $c=0.9$ and $\beta=1/2$ for the continuous controllers and call this rule the \textit{tempered selector}.

\begin{table*}[t]
\caption{Controller and estimator design values used throughout. Detection and correction measurement strengths, amplitudes and thresholds are trained per family and reported with each benchmark.}
\label{tab:design-values}
\begin{ruledtabular}
\begin{tabular}{llll}
symbol & value & meaning & defined in \\
\hline
$Q$ & $\operatorname{diag}(200,50)$ & LQR state weight, per plane
  & Sec.~\ref{subsec:lqr} \\
$R$ & $10^{-5}$ & LQR input weight & Sec.~\ref{subsec:lqr} \\
$p_{\mathrm{op}}$ & $0.2$ & floor; shifted point $s^\circ\simeq0.98$
  & Sec.~\ref{subsec:lqr} \\
$x_{\mathrm{ref}}^{\mathrm{sh}}$ & $(0.97,0)$ per plane & shifted-design reference & Sec.~\ref{subsec:lqr} \\
$u_{\max}$ & $400$ & per-channel drive cap, fixed & Sec.~\ref{subsec:lqr} \\
$K_p$ & $\operatorname{diag}(200,50)$ & inversion proportional gain
  & App.~\ref{subsec:sector_dynamic_inversion} \\
$K_d$ & $\operatorname{diag}(k_d,k_d/6)$, $k_d=60$ & inversion derivative gain
  & App.~\ref{subsec:sector_dynamic_inversion} \\
$\lambda$ & $0.01$ & Tikhonov regularization
  & App.~\ref{subsec:sector_dynamic_inversion} \\
$\alpha_x,\,\alpha_d$ & $0.2$, $0.15$ & state and slope smoothing
  & App.~\ref{subsec:sector_dynamic_inversion} \\
$c,\,\beta$ & $0.9$, $1/2$ & tempered selector contrast, tempering
  & Sec.~\ref{subsec:imm_approach} \\
$\theta$ & $0.99$ (3q), $0.995$ (5q) & calibrated pulse confidence, trained
  & Sec.~\ref{sec:shutdown} \\
 & $0.10$, $0.05$ & min.\ nontrivial mass, decision margin & Sec.~\ref{sec:shutdown} \\
$h$ & $10^{-4}$ (3q), $2\times10^{-4}$ (5q) & controller decision interval
  & Sec.~\ref{sec:benchmarks} \\
\end{tabular}
\end{ruledtabular}
\end{table*}
\paragraph*{Calibrated syndrome filter.}
When no Hamiltonian acts, commuting stabilizer observations close exactly on the class probabilities $p_a=\Tr(\Pi_a\rho_c)$ [Eq.~\eqref{eq:exact-syndrome-filter} of Sec.~\ref{sec:shutdown}]. If every nontrivial class is reached by exactly one correctable error at rate $\gamma$, the exact prediction and the normalized likelihood update over a step $\delta t$ are
\begin{equation}
\begin{split}
p_a^- &= \tfrac{1}{C}+\Bigl(p_a-\tfrac{1}{C}\Bigr)e^{-C\gamma\,\delta t},\\
p_a^+ &\propto p_a^-\exp\Bigl(2\sum_j\sqrt{\eta\kappa_j}\,\mu_{a,j}\,\Delta Y_j\Bigr),
\end{split}
\label{eq:calibrated-filter}
\end{equation}
with $C=2^{\,n-k}$ classes and the sum over the monitored channels; the common Gaussian factor cancels. This filter has no tempering or contrast reduction, starts from $p_a(0)=\delta_{a0}$, and is the estimator of the committed-recovery controller of Sec.~\ref{sec:shutdown}.

\paragraph*{The three-qubit instance.}
The four classes have templates
\begin{equation}
\begin{aligned}
\mu_0&=(+1,+1), & \mu_1&=(-1,+1), \\
\mu_2&=(-1,-1), & \mu_3&=(+1,-1)
\end{aligned}
\label{eq:3q-templates}
\end{equation}
for no error, $X_1$, $X_2$ and $X_3$. The syndrome graph of Eq.~\eqref{eq:hazard} is a two-bit hypercube in which every class has three neighbours and the escape rate is $3\gamma$.

\subsection{Control synthesis}
\label{subsec:lqr}
The sector models so far are exact, but for synthesis we simplify via a local linear surrogate. Writing a sector as $\dot x = Ax + uBx$ and linearizing about $(x^{\circ},u^{\circ}=0)$ gives
\begin{equation}
\delta\dot{x}=A\,\delta x+(Bx^{\circ})\,\delta u ,
\label{eq:linearization}
\end{equation}
so the only design question is where to evaluate the input direction $b_{\mathrm{lin}}=Bx^{\circ}$.

\paragraph*{The codespace degeneracy.}
For $w$ planes with signs $c_j$ and a common operating point $(s^\circ, p^\circ)$,
\begin{equation}
b_{\mathrm{lin}}=Bx^{\circ}
=\bigl(c_1p^{\circ},\,-c_1s^{\circ},\;\ldots,\;
        c_wp^{\circ},\,-c_ws^{\circ}\bigr)^{\!\top}.
\label{eq:b-codespace}
\end{equation}
At the codespace point $(s^{\circ},p^{\circ})=(1,0)$ the stabilizer entries vanish, and the syndrome coordinates are uncontrollable in the surrogate. Geometrically, the control rotates each plane along $s_j=\cos\theta$, $p_j=-\sin\theta$, and $ds_j/d\theta=0$ at $\theta=0$. At a \textit{shifted operating point}
\begin{equation}
p^{\circ}=p_{\mathrm{op}}>0,
\qquad
s^{\circ}=\sqrt{1-p_{\mathrm{op}}^2},
\label{eq:shifted-op}
\end{equation}
the stabilizer entries are nonzero,
\begin{equation}
b_{\mathrm{lin}}
=\bigl(c_1p_{\mathrm{op}},\,-c_1s^{\circ},\;\ldots,\;
        c_wp_{\mathrm{op}},\,-c_ws^{\circ}\bigr)^{\!\top}.
\label{eq:b-shifted}
\end{equation}
The bilinear plant~\eqref{eq:sector-bilinear} is therefore controllable everywhere except on the set $p=0$, which contains the codespace point. Restoring the stabilizer entry is necessary but not sufficient. For one plane,
\begin{equation}
\det\bigl[\,b_{\mathrm{lin}},\;Ab_{\mathrm{lin}}\,\bigr]
=2\kappa w\,c_j^{2}\,p_{\mathrm{op}}\sqrt{1-p_{\mathrm{op}}^{2}} ,
\label{eq:ctrb-det}
\end{equation}
which vanishes at $p_{\mathrm{op}}=0$ and also at $\kappa=0$, where the drift is isotropic and no operating point helps. The monitoring that causes the deadlock of Sec.~\ref{sec:deadlock} is also what makes the shifted surrogate controllable. Appendix~\ref{app:polar} treats $w>1$ and explains why stabilizability adds nothing here.

\paragraph*{Gains.}
For each sector we solve the continuous-time LQR problem~\cite{anderson_optimal_1990} with cost $J=\int_0^{\infty}\qty[\delta x^{\top}Q\,\delta x+R\,(\delta u)^2]dt$, $Q\succeq0$, $R>0$, through the algebraic Riccati equation
\begin{equation}
0={}A^{\top}P+PA+Q -P b_{\mathrm{lin}}R^{-1}b_{\mathrm{lin}}^{\top}P,
\label{eq:Ricatti}
\end{equation}
and set $K=R^{-1}b_{\mathrm{lin}}^{\top}P$. With $b_{\mathrm{lin}}$ at the codespace point, the stabilizer coordinates are undriven and open-loop stable, so the stabilizing solution gives them exactly zero gain. All partners carry the same damping, so the gain is uniform across planes and the drive is
\begin{equation}
u=k_p\sum_{j=1}^{w}c_j\,\hat p_j ,
\qquad k_p>0 .
\label{eq:codespace-gain}
\end{equation}
The syndrome error never enters this law, however large it is.

\paragraph*{Three LQR laws.}
\textit{Codespace LQR} applies
\begin{equation}
u_i(t)=\mathrm{sat}_{[-u_{\max},u_{\max}]}
\qty(-K_i\bigl[\hat{x}^{(i)}(t)-x_{\mathrm{ref}}^{(i)}\bigr]),
\label{eq:sat-control}
\end{equation}
with $x_{\mathrm{ref}}^{(i)}$ the codespace point of the sector and a per-channel bound $u_{\max}=400$ that does not depend on $\kappa$. \textit{Floored LQR} applies the same gain to an estimate whose partner coordinates are clamped,
\begin{equation}
\hat{p}_{\mathrm{op}}
=\operatorname{sgn}^{+}\!(\hat{p})\,
\max\bigl(|\hat{p}|,\,p_{\mathrm{op}}\bigr),
\qquad
\operatorname{sgn}^{+}\!(0)=+1,
\label{eq:floor}
\end{equation}
which guarantees $|u_i|\ge k_p\,p_{\mathrm{op}}\bigl|\sum_jc_j\bigr|$ whenever a nontrivial sector is active. \textit{Shifted-design LQR} instead solves Eq.~\eqref{eq:Ricatti} with $b_{\mathrm{lin}}$ at the shifted point~\eqref{eq:shifted-op}, $p_{\mathrm{op}}=0.2$ and $s^\circ\simeq0.98$, and applies the resulting gain to the raw estimate with stabilizer reference $0.97$ and no floor,
\begin{equation}
u_i(t)=\mathrm{sat}_{[-u_{\max},u_{\max}]}
\qty(-K_i^{\mathrm{sh}}\bigl[\hat{x}^{(i)}(t)-x_{\mathrm{ref}}^{\mathrm{sh}}\bigr]).
\label{eq:shifted-law}
\end{equation}
Because the shifted pair is controllable [Eq.~\eqref{eq:ctrb-det}], $K_i^{\mathrm{sh}}$ has a nonzero stabilizer component $K_s$. At the post-error point $(\hat s,\hat p)=(-1,0)$ the command $-K_s(-1-0.97)$ saturates at the cap in every sector of both codes.

\paragraph*{Regularized dynamic inversion.}
As an alternative, we use a law that prescribes a closed-loop velocity and realizes it in a regularized least-squares sense~\cite{enns_dynamic_1994}. It shares the stall geometry of codespace LQR, so the benchmarks use its operating-point variant (Appendix~\ref{subsec:sector_dynamic_inversion}).

\paragraph*{The three-qubit instance.}
The references are $x_{\mathrm{ref}}^{(1)}=x_{\mathrm{ref}}^{(3)}=(1,0)$ and $x_{\mathrm{ref}}^{(2)}=(1,0,1,0)$. With $c_j=+1$, the drive of Eq.~\eqref{eq:codespace-gain} is $k_p\hat p$ in the $w=1$ sectors and $k_p(\hat p_2+\hat p_3)$ in sector 2. Design values are collected in Table~\ref{tab:design-values}.

\section{The measurement-strength dilemma and switched-$\kappa$ control}
\label{sec:tradeoff}

The sector model~\eqref{eq:sector-general} explains how closed-loop performance depends on $\kappa$. We abbreviate its two rates as
\begin{equation}
a_s := a_s^{\rm noise},
\qquad
a_p := a_p^{\rm noise} + 2\kappa w ,
\label{eq:sector-rates}
\end{equation}
so that one plane obeys $\dot s_j = -a_s s_j + u\,p_j$ and $\dot p_j = -a_p p_j - u\,s_j$. The feedback moves the stabilizer only through the rotational partner, and without control the partner decays as
\begin{equation}
p_j(t) = p_j(0)\, e^{-a_p t}.
\label{eq:p-decay}
\end{equation}
At strong measurement, the control lever is therefore suppressed by the monitoring, and more so in sectors of higher multiplicity. In the bit-flip code, $a_s=4\gamma$ and $a_p = 4\gamma+2\kappa$ in sectors 1 and 3. At $\kappa = 400 \gamma$ the lever keeps $0.92$ of its amplitude over one controller macro-step and has dropped by an order of magnitude after thirty; at $\kappa = 50\gamma$ it keeps $0.99$ per step [Fig.~\ref{fig:lqr-mechanism}(c)]. The effect is Zeno-like~\cite{facchi_quantum_2002}: the measurement does not freeze the state, but it strongly damps the partner through which the controller steers the stabilizer.

Whether this costs control authority depends on how the drive is bounded. Adiabatic elimination of $p_j$ at large $\kappa$ gives
\begin{equation}
p_j \simeq -\frac{u\,s_j}{a_p},
\qquad
\dot{s}_j \simeq -\Bigl(a_s + \frac{u^{2}}{a_p}\Bigr) s_j ,
\label{eq:adiabatic}
\end{equation}
so a control of amplitude $u$ depletes a faulty sector at the effective rate $u^2/a_p$. All controllers share the fixed bound $|u_q|\le u_{\max}=400$, so the achievable corrective rate
\begin{equation}
\Lambda(\kappa) = \frac{u_{\max}^{2}}{a_p^{\rm noise} + 2\kappa w}
\;\xrightarrow{\;\kappa \gg \gamma\;}\; \frac{u_{\max}^{2}}{2\kappa w}
\label{eq:Lambda}
\end{equation}
decreases with the measurement strength, faster in sectors of higher multiplicity. When the drive cannot grow with the monitoring, strong measurement removes corrective authority.

The conditional dynamics add a stochastic cost. The innovation term of the filter contains
\begin{equation}
ds_{j} \supset 2\sqrt{\eta\kappa}\,\bigl(1 - s_{j}^{2}\bigr)\,
dW_{j},
\label{eq:backaction}
\end{equation}
so back-action injects diffusion of amplitude $\sqrt{\kappa}\,(1 - s_{j}^{2})$ into the stabilizer coordinate. At $s_j=\pm1$ this term vanishes and strong monitoring only helps detection. A recovery, however, must cross the interior of the interval, where $1-s_j^2\sim1$, and is exposed to back-action for the whole excursion.

Both costs act only while a correction is under way. The measurement strength should therefore be large while the register idles and small during correction,
\begin{equation}
\kappa(t) =
\begin{cases}
\kappa_{\mathrm{det}}, & \hat{\sigma}(t) = 0,\\[2pt]
\kappa_{\mathrm{cor}}, & \hat{\sigma}(t) \neq 0,
\end{cases}
\label{eq:switched-kappa}
\end{equation}
where $\hat{\sigma}(t)$ is the active sector; in the implementation the switch is keyed to a nonzero commanded control. In the benchmarks $\kappa_{\mathrm{cor}}$ is trained; the selected LQR schedules use $\kappa_{\mathrm{det}}=400$ and $\kappa_{\mathrm{cor}}=50$. Lowering the measurement during correction extends the partner lifetime by $(4\gamma+2\kappa_{\mathrm{det}}w)/(4\gamma+2\kappa_{\mathrm{cor}}w)$, restores $\Lambda$, and reduces the back-action along the recovery path.

\subsection{The information cost of weakening the measurement}
\label{subsec:info-cost}
The same parameter sets how fast the record informs the estimator. Integrating $dY_j$ over a window $\tau$ separates $s_j=\pm1$ by a mean displacement $4\sqrt{\eta\kappa}\,\tau$ against a noise standard deviation $\sqrt{\tau}$, so the single-generator discrimination error is
\begin{equation}
\varepsilon_j=\Phi\!\left(-2\sqrt{\eta\kappa\,\tau}\right),
\label{eq:syndrome-snr}
\end{equation}

\subsection{Which controllers benefit}
\label{subsec:switch-scope}
The benefit grows with $w$, because the partner damping is $2\kappa w$. The benchmarks follow this ordering. For floored LQR, reducing the measurement during correction gains $0.0957$ in $F_e(3)$ when all three parities of the three-qubit code are monitored ($w=2$ in every sector; two-contrast adjusted 95\% interval $[0.0579,0.1334]$), $0.0071$ with only the two generators ($w=1$ in the outer sectors; $[-0.0378,0.0521]$), and $0.0998$ in the five-qubit code with $w$ up to 4 (ordinary interval $[0.0030,0.1967]$, adjusted $[-0.0110,0.2106]$). The limit $\kappa_{\mathrm{cor}}\to0$ removes both the partner damping and the back-action during the rotation; it is the committed recovery of Sec.~\ref{sec:shutdown}.

\section{A controllability degeneracy of the reduced model}
\label{sec:deadlock}

Section~\ref{sec:tradeoff} identified a stochastic, plant-side cost of strong monitoring. The reduced model also exposes a deterministic failure on the estimator--controller side: codespace-linearized LQR can impose zero control before any correction begins. This is a statement about where the bilinear control map $x\mapsto Bx$ loses rank, and the shifted design of Sec.~\ref{subsec:lqr} removes it.

\subsection{The $p=0$ deadlock}

At the codespace point, $b_{\mathrm{lin}}=(0,-1)^{\top}$ and $A_i$ is diagonal, so $\delta s$ is uncontrollable in the surrogate. An uncontrollable, open-loop-stable mode receives zero Riccati gain, which gives the gain structure~\eqref{eq:codespace-gain}, $u=k_p\hat p$ for a single plane. In the reduced filter, every drift and innovation term of the partner block is proportional to the partner coordinates themselves or to the control [Eqs.~\eqref{eq:9d-damping}--\eqref{eq:closed-filter}]. With $u=0$ the subspace $\hat p=\hat q=0$ is exactly invariant, and measurement noise alone never regenerates a lever. The point $(\hat s,\hat p,u)=(-1,0,0)$ is therefore a fixed point of the estimator--controller cascade at every $\kappa$, and in exact arithmetic the codespace-linearized law never fires.

\subsection{Coupled innovations and the local escape exponent}
Weak and strong measurement differ in how the zero-partner set responds to a perturbation. A residual seed of $\hat p$ commands $u=k_p\hat p$, which regrows the lever through $-u\hat s$. For a frozen post-$X_1$ background $(s_{12},s_{23},s_{13})=(-1,+1,-1)$ in the two-current model, the equations for $p=\langle YZI\rangle$ and $q=\langle YIZ\rangle$ are coupled. With $u=Gp$ to linear order,
\begin{align}
 dp={}&[-(4\gamma+2\kappa)p+Gp]dt\nonumber\\
 &+2\sqrt{\eta\kappa}[p\,dW_{12}+(q-p)dW_{23}],\\
 dq={}&[-(4\gamma+2\kappa)q+Gp]dt\nonumber\\
 &+2\sqrt{\eta\kappa}[q\,dW_{12}+(p-q)dW_{23}].
 \label{eq:phat-lin}
\end{align}
The symmetric mode $p=q$ is invariant, and the unflipped-channel innovation partially cancels it; treating that channel as an independent $-p\,dW_{23}$ noise would double-count a fluctuation and give the wrong critical strength. For a symmetric mode with aggregate gain $G$, physical damping $a$, and $w$ flipped measured channels, the local multiplicative-noise equation and its It\^o log-growth rate~\cite{khasminskii_stochastic_2012} are
\begin{align}
 dz&=(G-a-2w\kappa)z\,dt
 +2\sqrt{\eta\kappa}\,z\sum_{j=1}^w dW_j,\nonumber\\
 \lambda(\kappa)&=G-a-2w\kappa(1+\eta),
 \label{eq:escape-exponent}
\end{align}
with $G=wk_p$ for $w$ equal planes driven symmetrically. The three terms are the controller's regrowth, the measurement-enhanced damping, and the sample-path contraction of the multiplicative innovation. For $\lambda>0$, a seed is amplified; for $\lambda<0$, ignition attempts collapse back to the stall. The exponent describes a small nonzero seed in the frozen model. It cannot ignite the exactly zero solution and is not a global fidelity boundary.

\begin{figure*}[tp]
\includegraphics[width=\textwidth]{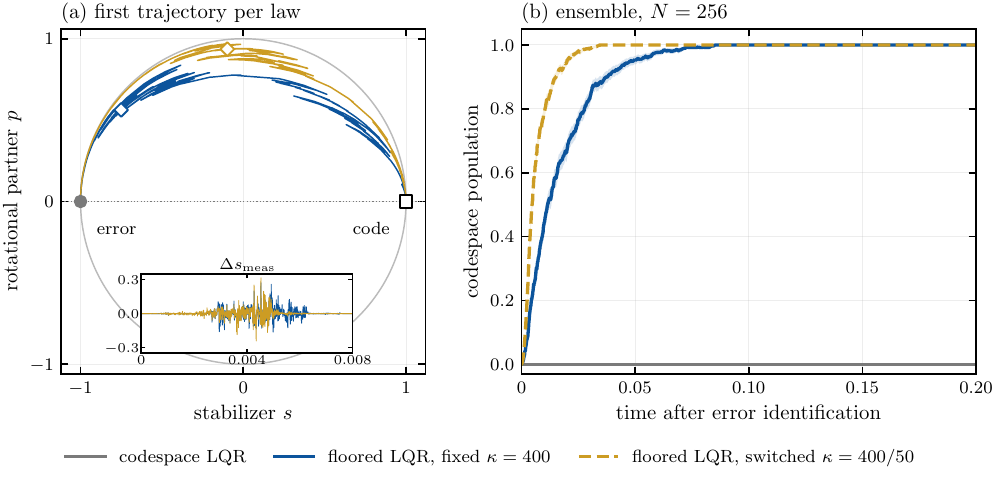}
\caption{Conditional recovery of an isolated, already identified $X_1$ error in the three-qubit code, with further physical errors disabled ($\gamma_{\rm rec}=0$) and measurement innovations active at efficiency $\eta=0.75$. The fixed-channel filter reduces exactly to $(s,p)$ in the occupied block, and its syndrome probabilities gate the drive. (a) First generated trajectory for each law, without postselection, sampled at time interval $10^{-5}$; diamonds mark $t=0.004$; square marks final time. Inset: measurement-induced stabilizer update $\Delta s_{\rm meas}$ for these records as a function of time. (b) Mean codespace population over 256 trajectories per law ($\pm1$ SEM), to $T=0.2$. Codespace LQR stays deadlocked; both floored laws return to the codespace, with all 256 trajectories above population 0.99 at $T$. Fixed measurement uses $\kappa=400$; switched measurement uses 50 during actuation and 400 otherwise. Cap 400, floor 0.2, hold $h=10^{-4}$, integration step $10^{-5}$, gains designed at $\gamma_{\rm design}=1$.}
\label{fig:phaseplane}
\end{figure*}

For the design values of Table~\ref{tab:design-values}, with $a=4\gamma$ and $w=1$, $k_p(\kappa)$ decreases slowly with $\kappa$ while the losses grow linearly, and $\lambda$ crosses zero at
\begin{equation}
\kappa_c\simeq3.9\times10^{2}\,\gamma\quad(\eta=1),
\label{eq:kappa-c}
\end{equation}
and at $4.4\times10^{2}\gamma$ and $4.9\times10^{2}\gamma$ for $\eta=0.7$ and $0.5$. The controller holds each command for a decision interval $h$, and the sampled law has its own exponent: at $h=10^{-4}$ and $\kappa=400$, continuous feedback predicts $\lambda=-31.8$ while the sample-and-hold calculation gives $25.4\pm0.5$, with zero crossings at $394$ and $405$. The deadlock is therefore marginal at the strongest monitoring used in the benchmarks and is not their dominant failure mode.

The dynamic-inversion law has the same geometry. At $\hat p=0$ its controllable direction $g(\hat x)=(0,-\hat s)^{\top}$ is orthogonal to the desired stabilizer restoration, and the projected control of Eq.~\eqref{eq:ureg_full} vanishes. Regularization bounds the gain but cannot rotate the projection; the operating-point variant $g_{\mathrm{eff}}=B\,x^{\circ}$ removes the degeneracy.

\subsection{Remedies}
The bilinear plant is controllable at every point off $p=0$ [Eq.~\eqref{eq:b-shifted}], so the deadlock comes from where the surrogate is evaluated. The shifted design fixes this at the synthesis stage: the reachability condition~\eqref{eq:ctrb-det} holds so that the gain acquires a stabilizer component, and the command $-K_s(\hat s-0.97)$ is nonzero at the post-error point. The floor~\eqref{eq:floor} keeps the codespace gain and evaluates it on a clamped partner, which gives
\begin{equation}
u=k_p\,p_{\mathrm{op}}\neq 0
\label{eq:ignition}
\end{equation}
at the deadlock point. Neither is a retuning of $Q$ and $R$, and neither acts when the sector estimator reports the codespace. In the benchmarks the two are interchangeable: adding the floor to the shifted design changes the terminal fidelity by $-0.0010$, $-0.0037$, and $-0.0031$ in the three-qubit comparisons with two and three monitored currents and in the five-qubit comparison, each with a 95\% interval containing zero (intervals $[-0.0160,0.0140]$, $[-0.0154,0.0080]$, and $[-0.0651,0.0590]$). We therefore use the shifted design, whose ignition follows from the geometry of the model without an extra clamp.

\begin{figure*}[tp]
\includegraphics[width=0.85\textwidth]{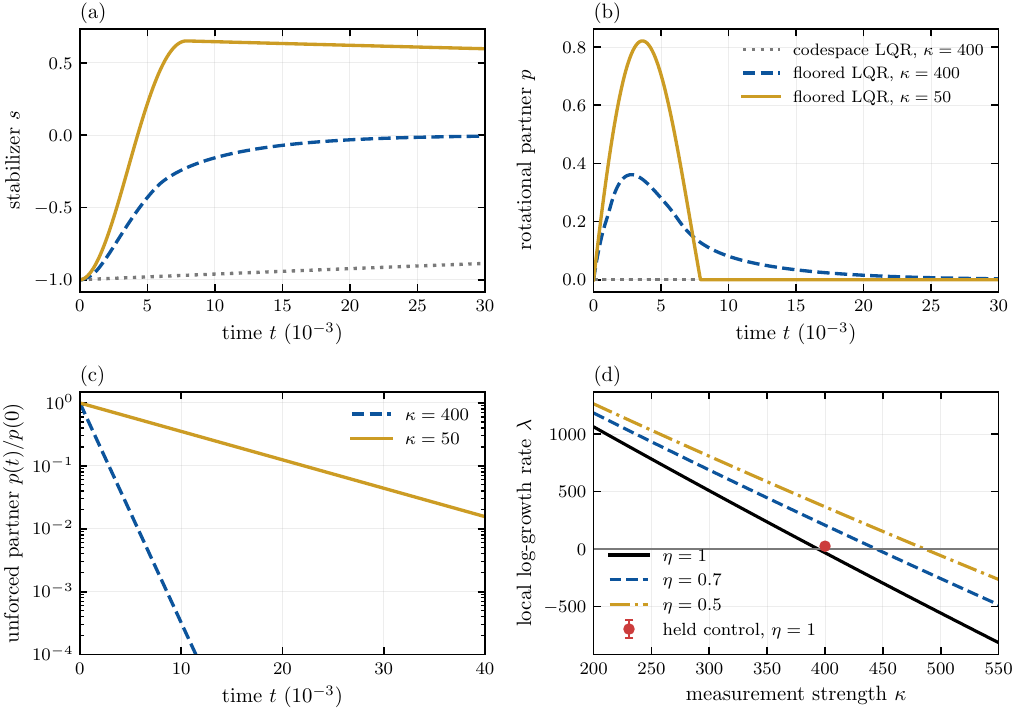}
\caption{Bilinear mechanism and deadlock analysis. (a,b) Deterministic one-plane drift at $\gamma=1$, cap 400, and the codespace Riccati gain designed at each $\kappa$, from $(-1,0)$ for $\kappa=400$ and $\kappa=50$. The floor ($p_{\rm floor}=0.2$) starts recovery; baseline LQR keeps a zero partner. These are drift illustrations, not stochastic trajectories. (c) Partner lifetime under fixed and reduced measurement, one flipped channel. (d) Frozen-background log-growth rates under continuous feedback; the red point is the sample-and-hold value at $\eta=1$, $h=10^{-4}$.}
\label{fig:lqr-mechanism}
\end{figure*}

\subsection{Conditional recovery after an identified error}

Fig.~\ref{fig:phaseplane} tests the return to the codespace after a single identified error with no later errors. In the occupied code/$X_1$ block, $s=\langle ZZI\rangle=\langle ZIZ\rangle$, $p=\langle YZI\rangle=\langle YIZ\rangle$, $\langle IZZ\rangle=1$, and the conditional equations reduce exactly to
\begin{align}
 ds&=up\,dt+2\sqrt{\eta\kappa}(1-s^2)dW,\\
 dp&=(-us-2\kappa p)dt-2\sqrt{\eta\kappa}sp\,dW.
 \label{eq:isolated-recovery}
\end{align}
The drive uses these moments and the syndrome probabilities $\pi_0=(1+s)/2$, $\pi_1=(1-s)/2$, with decision margin 0.05. The floor starts the motion from $(-1,0)$ and the selector switches it off near the codespace, where $(1,0)$ is absorbing; without the floor, $(-1,0)$ remains an equilibrium. The filter agrees with an independent register-plus-reference density replay to below $7\times10^{-14}$. Stronger monitoring damps $p$ and slows the transfer term $up$ even when the path lies further inside the plane, which is why the equal-time markers matter; single realizations in panel (a) need not reproduce the ensemble timing of panel (b). A repeat at half the integration step gives the same outcome. With ongoing errors, returns to the codespace are interrupted, and the net benefit is assessed by the memory benchmarks.

The measurement conditioning is essential here. Without innovations, the deterministic drift obeys
\begin{equation}
 \frac{d}{dt}(s^2+p^2)=-8\gamma(s^2+p^2)-4\kappa p^2,
 \label{eq:drift-radius-decay}
\end{equation}
which for $\gamma>0$ contracts to the origin and cannot show recovery to the codespace. Fig.~\ref{fig:lqr-mechanism} keeps a short drift illustration of ignition and damping, together with the escape analysis.

The degeneracy of Eq.~\eqref{eq:b-codespace} is exact and code independent. Its observable consequence requires a negative escape exponent~\eqref{eq:escape-exponent} and a register that lingers near the stalled point. The isolated-error diagnostic of Fig.~\ref{fig:phaseplane} shows the deadlock directly; in the noisy-memory benchmarks it is one of several failure modes, and codespace LQR without a floor coincides with no feedback.

\begin{figure*}[tp]
\includegraphics[width=0.9\textwidth]{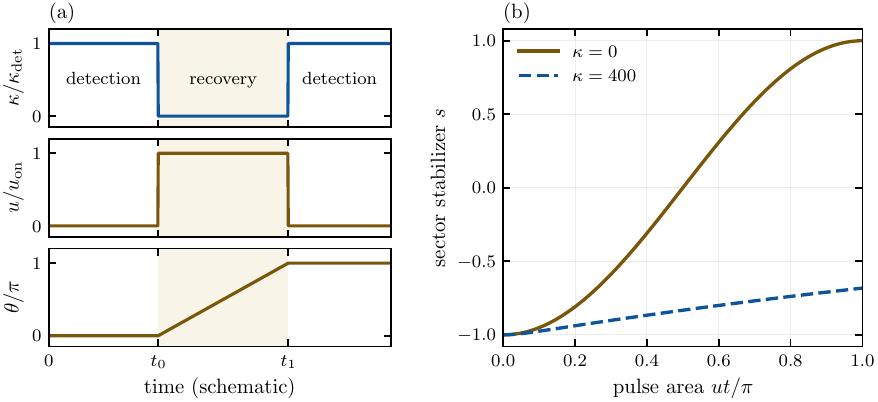}
\caption{Recovery policy and mechanism. (a) Detection, measurement-off rotation between $t_0$ and $t_1$, and return to detection. (b) Deterministic sector drift with two measured channels anticommuting with the control, $u=200$, and no physical noise. Continuous measurement damps the partner and prevents a nominal $\pi$ pulse from completing the stabilizer transfer. The curves illustrate Eq.~\eqref{eq:sector-general} and are not ensemble data.}
\label{fig:mechanism}
\end{figure*}

\section{Committed recovery with measurement shutdown}
\label{sec:shutdown}
We now consider the case where the measurement is completely shut off during recovery. Let $\pi_a$ be the estimator's probability for syndrome class $a$, with $a=0$ the codespace. While idle, the controller measures at $\kappa_{\rm det}$ and tracks the most likely nontrivial class. It triggers when $\pi_a\ge\theta$, the nontrivial mass is at least 0.10, and $\pi_a>\pi_0+0.05$, and then commits to the corresponding Pauli rotation with
\begin{equation}
 \kappa_j(t)=0,\qquad u_a(t)=u_{\rm on},\qquad
 \int_{\rm pulse}u_a(t)dt=\pi .
 \label{eq:pulse}
\end{equation}
The last held command is shortened if needed to complete the area exactly on the decision grid, and all other controls are zero during the pulse. Measurement resumes at completion, while physical noise never stops. 

The measurement-on ablation keeps the same trigger and area rule but measures at $\kappa_{\rm det}$ throughout. The drive cap is the same $|u_q|\le400$ as for every other controller, so switching off the measurement neither removes nor increases the drive allowance. With $u_{\rm on}=400$ the blind interval is $\tau_p=h\lceil\pi/(u_{\rm on}h)\rceil=0.008$ in the five-qubit tests, and the probability of at least one physical jump during it is $1-e^{-5\tau_p}\simeq3.92\%$. Faster inference cannot remove this cost. Fig.~\ref{fig:mechanism} illustrates the mechanism with amplitude 200, the value selected in the three-qubit benchmarks (Table~\ref{tab:three-unified}).

\subsection{An exact reduced update at shutdown boundaries}
Let $\Pi_a$ project onto syndrome class $a\in\mathbb F_2^r$ and $p_a=\operatorname{Tr}(\Pi_a\rho_c)$. A Pauli $E_\mu$ shifts the syndrome by $s(E_\mu)$. Define the Markov generator and the permutation
\begin{equation}
 (Qp)_a=\sum_\mu\gamma_\mu[p_{a\oplus s(E_\mu)}-p_a],
 \qquad (M_Pp)_a=p_{a\oplus s(P)}.
 \label{eq:syndrome-generator}
\end{equation}
During detection, with zero Hamiltonian, commuting stabilizer observations close on $p$:
\begin{align}
 dp_a={}&(Qp)_a dt+2p_a\sum_j\sqrt{\eta_j\kappa_j}
 (s_{ja}-\bar s_j)dW_j,\nonumber\\
 \bar s_j&=\sum_a s_{ja}p_a.
 \label{eq:exact-syndrome-filter}
\end{align}
This is the classical error-tracking filter of Ref.~\cite{vanhandel2005}. The question here is whether a finite coherent recovery can be inserted without propagating its intermediate partners.

\begin{theorem}[Exact shutdown-boundary reduction]
\label{thm:shutdown}
Consider alternating detection intervals governed by Eq.~\eqref{eq:exact-syndrome-filter} and committed recovery intervals $[t_0,t_1]$. On each recovery interval, assume: (i) all stabilizer measurements are off; (ii) $H_c(t)=u(t)P/2$ with one fixed Hermitian Pauli and $\int_{t_0}^{t_1}u(t)dt=\pi$; (iii) the Pauli-noise generator $\mathcal L_N$ is constant and commutes with $\mathcal H_P(\rho)=-i[P/2,\rho]$. The axis, duration, and pulse profile are fixed by the record before the pulse.

\begin{table*}[t]
\caption{Frozen settings and held-out three-qubit endpoints at $T=3$ ($N=512$ per method, $\eta=0.75$, $\gamma=1$, monitored parities $Z_1Z_2$, $Z_2Z_3$, and $Z_1Z_3$). $\theta$ is the posterior-confidence threshold of the pulse rules and $\epsilon$ the local-$Z$ threshold of the Borah rule; a dash denotes the common continuous-controller selector. $B_\kappa$ and $A_u$ are mean measurement expenditure and drive area.}
\label{tab:three-unified}
\begin{ruledtabular}
\begin{tabular}{lrrrrllrr}
Method & $\kappa_{\rm det}$ & $\kappa_{\rm cor}$ & $u_{\max}^{\rm used}$ & $\theta$/$\epsilon$ & $F_e(3)$ & $F(3)$ & $B_\kappa$ & $A_u$\\
Shutdown, calibrated & 400 & 0 & 200 & 0.99 & $0.9473\pm0.0099$ & $0.9117\pm0.0121$ & 3432 & 27.8\\
Shutdown, tempered & 400 & 0 & 200 & 0.95 & $0.9258\pm0.0116$ & $0.8841\pm0.0137$ & 3436 & 27.2\\
Floored LQR, fixed & 100 & 100 & 400 & -- & $0.7969\pm0.0178$ & $0.7563\pm0.0188$ & 900 & 47.3\\
Floored LQR, switched & 400 & 50 & 400 & -- & $0.8926\pm0.0137$ & $0.8574\pm0.0153$ & 3459 & 47.8\\
Shifted-design LQR & 400 & 50 & 400 & -- & $0.8638\pm0.0151$ & $0.8217\pm0.0166$ & 3437 & 48.7\\
Dynamic inverse & 100 & 100 & 400 & -- & $0.7046\pm0.0202$ & $0.6315\pm0.0211$ & 900 & 91.6\\
Ahn & 400 & 400 & 400 & -- & $0.6863\pm0.0196$ & $0.6376\pm0.0204$ & 3600 & 3171.1\\
Geremia first order & 400 & 400 & 100 & -- & $0.4801\pm0.0212$ & $0.2804\pm0.0193$ & 3600 & 813.2\\
Borah (published) & 400 & 400 & 400 & 1.05 & $0.9199\pm0.0120$ & $0.8426\pm0.0160$ & 3600 & 79.7\\
Measurement-on pulse & 400 & 400 & 200 & 0.95 & $0.5251\pm0.0221$ & $0.2474\pm0.0190$ & 3600 & 321.9\\
No feedback / codespace LQR & 400 & 400 & 400 & -- & $0.5039\pm0.0221$ & $0.1309\pm0.0149$ & 3600 & 0.0\\
\end{tabular}
\end{ruledtabular}
\end{table*}

Then, with $\tau=t_1-t_0$, the conditional channel averaged over unobserved noise during recovery is
\begin{equation}
 \mathcal E_{\rm off}=\operatorname{Ad}_P\circ e^{\tau\mathcal L_N},
 \qquad p(t_1)=M_Pe^{\tau Q}p(t_0),
 \label{eq:shutdown-exact}
\end{equation}
where $\operatorname{Ad}_P(\rho)=P\rho P$. A correctly initialized syndrome filter is therefore exact throughout detection and at every completed recovery boundary, even when later pulses depend on earlier records and the register is entangled with an untouched reference.
\end{theorem}
The proof is in Appendix~\ref{app:shutdown-proof}. The assumptions hold for bit-flip noise with $X_i$ recovery and for independent isotropic depolarization with single-qubit Pauli recovery, the two models used here: equal rates for the three Pauli axes at each site make the local channel covariant under single-qubit rotations. Anisotropic Pauli noise or multi-qubit control need not satisfy them; Appendix~\ref{app:shutdown-proof} bounds the error in that case.

Coherent partners are thus needed during a pulse but not to initialize the next detection interval. The known-control update is the permutation
\begin{equation}
 p_a\longleftarrow p_{a\oplus s(P)},
 \label{eq:completion}
\end{equation}
applied at physical completion, while the noise prediction runs throughout the blind interval. Applying it when the pulse is requested would be wrong in the presence of latency. In an earlier 4096-trajectory ablation with the tempered five-qubit posterior (Supplemental Material), adding this completion update raised $F_e(T)$ from $0.5672\pm0.0077$ to $0.6779\pm0.0073$.

The theorem gives exact syndrome information for this restricted architecture. It does not reconstruct the logical state from syndrome probabilities, since logically distinct errors can share a syndrome. The main shutdown controller uses the calibrated filter~\eqref{eq:calibrated-filter}; the tempered selector is kept as an ablation and is not made exact by the theorem. Residual measurement during a pulse lies outside the theorem and is tested in Sec.~\ref{sec:native-robustness}. Interrupted monitoring itself is not new: earlier CQEC work discusses postponing recovery until the measurement stops~\cite{chase_efficient_2008, martin2015hybrid}, and pausing the meter drive before a correction has been proposed to limit measurement-induced dephasing~\cite{halaski_obstacles_2026}. Our contribution is the boundary identity, its assumptions, and its use to connect bounded recovery with a reduced estimator.

\section{Benchmarks, metrics, and fairness}
\label{sec:benchmarks}
\subsection{Logical information and terminal recovery}
Our primary metric is the entanglement fidelity~\cite{schumacher_sending_1996} of an encoded qubit initially maximally entangled with an untouched reference,
\begin{equation}
 |\Phi_L\rangle=\frac{|0_L0_R\rangle+|1_L1_R\rangle}{\sqrt2}.
\end{equation}
Let $\mathcal E_t$ include physical noise, measurement, and online feedback. At each output time we apply a common ideal syndrome recovery $\R(\rho)=\sum_a E_a\Pi_a\rho\Pi_a E_a^\dagger$ with fixed minimum-weight representatives $E_a$, followed by decoding:
\begin{equation}
 F_e(t)=\langle\Phi|[(\mathcal V_{\rm dec}\circ\R\circ\mathcal E_t)
 \otimes\id_R](|\Phi_L\rangle\langle\Phi_L|)|\Phi\rangle.
 \label{eq:fe}
\end{equation}
This terminal recovery is part of the evaluation map only; it is not available to the online controller and does not reset the trajectory. 

The second metric is logical-basis survival before recovery,
\begin{equation}
 F(t)=\frac12\sum_{b=0}^1\langle b_L|\mathcal E_t(|b_L\rangle\langle b_L|)|b_L\rangle.
\end{equation}
In the logical-Pauli expansion of the encoded-Bell trajectory, $F_e=\sum_a|\psi_{a,I}|^2$ and $F=|\psi_{0,I}|^2+|\psi_{0,Z}|^2$. $F$ is neither the codespace population nor the average fidelity over all logical inputs, and the usual average-fidelity relation does not connect the two metrics.

\subsection{Common protocol}
Both codes are tested with horizon $T=3$, efficiency $\eta=0.75$, and zero added feedback latency. Every method is constrained by $\kappa_j\le400$ and $|u_q|\le400$ with $H=\sum_q u_qP_q/2$, and the drive cap stays fixed when the measurement changes. Simultaneous drives are allowed, while the sector-based laws drive one channel at a time, so this is a per-channel peak comparison and not an equal-power comparison. We record the integrated measurement expenditure and drive area,
\begin{equation}
 B_\kappa=\sum_{j}\int_0^T\kappa_j(t)dt,
 \qquad A_u=\sum_q\int_0^T|u_q(t)|dt,
 \label{eq:resources}
\end{equation}
which are reported but not matched. The register and reference are propagated as a normalized encoded-Bell amplitude vector. Physical Pauli errors are sampled from exponential waiting times, independently of the measurement randomness. Positive QND measurement maps and unitary control maps are split at the integration step, and errors within a step are applied at its boundary (Appendix~\ref{app:numerics}). All controllers share the same physical equations, output times, integration scheme, and terminal recovery, and none receives hidden error times, labels, or the physical state.

\subsection{Controllers and published comparators}
Codespace LQR without a floor is represented by the no-feedback trajectory, since its zero-partner state is invariant. Floored LQR is tested at fixed measurement and with reduced measurement during correction; shifted-design LQR (without a floor) and regularized dynamic inversion are further native-estimator controls. Gains are synthesized offline at the correction-phase measurement strength.

We also implement three published approaches. The hn et al. adaptation uses codespace-gradient feedback from a full auxiliary conditional state~\cite{ahn_continuous_2002}. The Geremia et al. adaptation uses the same gradient law with the first-order projected filter of Ref.~\cite{chase_efficient_2008}, which retains the union of the single-recovery branches and need not remain positive under simultaneous control; all syndrome-probability violations are recorded, with no trajectory rejection or repair. Both gradient laws use an explicit ignition rule when the gradient vanishes but an error sector is more likely than the codespace (Appendix~\ref{app:numerics}). The Borah et al. comparator propagates a full auxiliary density operator driven by the observed currents and the applied Hamiltonian~\cite{borah_measurement_2022}. These are specified implementations of the published approaches, not reproductions of their original figures.

\subsection{Tuning and statistics}
Each family is tuned independently on training trajectories. Within a family, the candidate with the highest training $F_e(3)$ among those with no failed trajectory and no syndrome-probability violation above $10^{-8}$ is frozen, then tested on held-out trajectories. Training grids are finite and training effort is not equalized across families, so the selected settings are not claimed to be global optima. All arms of a test share one seed schedule for physical errors and measurement noise, so policy differences are paired, although each controller produces its own measurement record. Quoted $\pm$ values are Monte Carlo standard errors, and decay bands are pointwise normal 95\% intervals clipped to $[0,1]$ for display. The seven shutdown-versus-family $F_e(3)$ contrasts receive Bonferroni-adjusted intervals, and the two switched-versus-fixed LQR contrasts form a separate family. Repeats at a finer integration step are sensitivity checks, not discretization-error bounds; an unresolved difference does not establish equivalence; and we infer no logical lifetime from the short, nonexponential curves. Validation scripts, full training records including failures, and all test trajectories are supplied, and earlier exploratory studies are collected in the online repository. 

\section{Unified three-qubit controller comparison}
\label{sec:three-unified}
We first apply the protocol to the bit-flip code of Sec.~\ref{sec:sector}, for which the reduced model and the deadlock analysis are written and the Borah et al. published detector is defined~\cite{borah_measurement_2022}. Each qubit flips at rate $\gamma=1$, so the register jump rate is 3. The decision interval is $h=10^{-4}$, the integration step $\delta t=10^{-5}$, and register plus reference form a 16-amplitude vector.

\begin{figure*}[tp]
\includegraphics[width=0.9\textwidth]{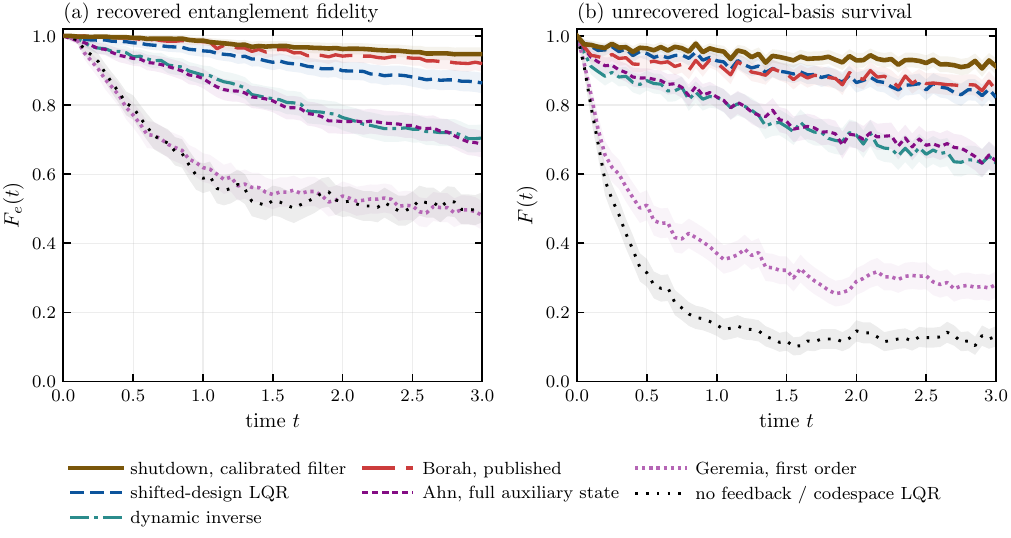}
\caption{Three-qubit benchmark with the three monitored parities of the published Borah protocol~\cite{borah_measurement_2022}. (a) Entanglement fidelity after ideal terminal recovery and decoding. (b) Unrecovered logical-basis survival. One representative per family: calibrated shutdown (Theorem~\ref{thm:shutdown}), shifted-design LQR without a floor, dynamic inversion, the published Borah protocol, Ahn protocol, Geremia protocol, and no feedback; the other arms are in Table~\ref{tab:three-unified}. 512 held-out trajectories per curve with paired error histories; bands are pointwise 95\% intervals. Peak constraints are common; integrated expenditure is not matched.}
\label{fig:three-unified}
\end{figure*}

\paragraph*{Measurement configuration and estimators.}
Every method monitors the three parities $Z_1Z_2$, $Z_2Z_3$, and $Z_1Z_3$, the configuration of the published Borah protocol. The third parity is the product of the two generators and adds no syndrome information, but it adds an independent noisy record of the same syndrome. Every $X_q$ then flips two monitored channels, so each sector has $w=2$ and partner damping $4\gamma+4\kappa$. The fixed-channel filter of Theorem~\ref{thm:closure} still has six coordinates, because $Z_1Z_3\in\mathcal{S}$. The continuous native controls use it with the sector projection and the tempered selector. Bit-flip noise commutes with $X_q$ recovery, so Theorem~\ref{thm:shutdown} applies, and the main shutdown uses the calibrated four-class filter~\eqref{eq:calibrated-filter} with $e^{-4\gamma\delta t}$, a likelihood over all three records, and the permutation at completion. Ahn's law propagates a 64-coefficient auxiliary state and Geremia's a ten-coefficient first-order union filter. The Borah estimator starts in $|000\rangle$ and drives qubit $q$ with amplitude $\lambda_q$ whenever $1-\langle Z_q\rangle$ exceeds a tolerance $\epsilon$, with simultaneous drives allowed; we implement this protocol literally. Appendix~\ref{app:three-two-gen} repeats the comparison with only the two generators monitored.

\paragraph*{Tuning and tests.}
The training grid has 69 candidates over the same families, with Ahn and Geremia amplitudes down to 50 and the Borah grid $\kappa\in\{100,200,400\}$, $\lambda\in\{200,400\}$, $\epsilon\in\{1.05,1.5,1.9\}$, each scored on 64 training trajectories at $\delta t=2\times10^{-5}$. Shifted-design LQR selects $\kappa_{\rm det}=400$, $\kappa_{\rm cor}=50$; calibrated shutdown selects amplitude 200 and $\theta=0.99$; the Borah protocol selects $\kappa=400$, $\lambda=400$, and $\epsilon=1.05$, the tolerance chosen in Ref.~\cite{borah_measurement_2022}. Every method is tested on 512 trajectories, and a refinement repeats every test at $\delta t=5\times10^{-6}$ with 256 independent trajectories.

\paragraph*{Results.}
Calibrated shutdown has the highest endpoints, $F_e(3)=0.9473\pm0.0099$ and $F(3)=0.9117\pm0.0121$ (Fig.~\ref{fig:three-unified}, Table~\ref{tab:three-unified}). Under identical monitoring it exceeds the published Borah protocol, $F_e(3)=0.9199\pm0.0120$, by $0.0273$, with ordinary 95\% interval $[0.0087,0.0460]$ and Bonferroni-adjusted interval $[0.0018,0.0529]$; the $F(3)$ gain is $0.0691$, interval $[0.0433,0.0949]$. It does so with slightly less measurement (3432 against 3600) and about a third of the drive area (27.8 against 79.7). Its margins over dynamic inversion, Ahn, and Geremia are 0.243, 0.261, and 0.467.

Among continuous laws, floored LQR with switched measurement comes closest, $F_e(3)=0.8926\pm0.0137$; shutdown exceeds it by $0.0547$, adjusted interval $[0.0221,0.0873]$. Shifted-design LQR reaches $F_e(3)=0.8638\pm0.0151$, and floored LQR at the same schedule is higher by $0.0288$, interval $[0.0031,0.0544]$. Reduced measurement during correction raises floored LQR by $0.0957$ over its fixed-measurement version (Sec.~\ref{subsec:switch-scope}), consistent with the stronger partner damping of the $w=2$ sectors. Measurement scheduling dominates the pulse comparison: switching the measurement off during the committed rotation raises the tempered pulse from $0.5251$ by $0.4007$, interval $[0.3515,0.4499]$, and replacing the tempered selector by the calibrated filter adds $0.0215$, interval $[0.0032,0.0398]$.

The refinement preserves the ordering: calibrated shutdown gives $0.9570\pm0.0127$ and the Borah protocol $0.9219\pm0.0168$, a paired difference of $0.0352$, interval $[0.0058,0.0646]$. No trajectory fails and no syndrome-probability violation above $10^{-8}$ occurs. Monitoring only the two generators lowers the leading methods, by 0.041 for calibrated shutdown and 0.104 for the Borah rule, and reduces the measurement expenditure by a third (Appendix~\ref{app:three-two-gen}); Sec.~\ref{sec:discussion} explains this trade-off.

\subsection{Results}
\label{sec:results}
\begin{figure*}[tp]
\includegraphics[width=0.9\textwidth]{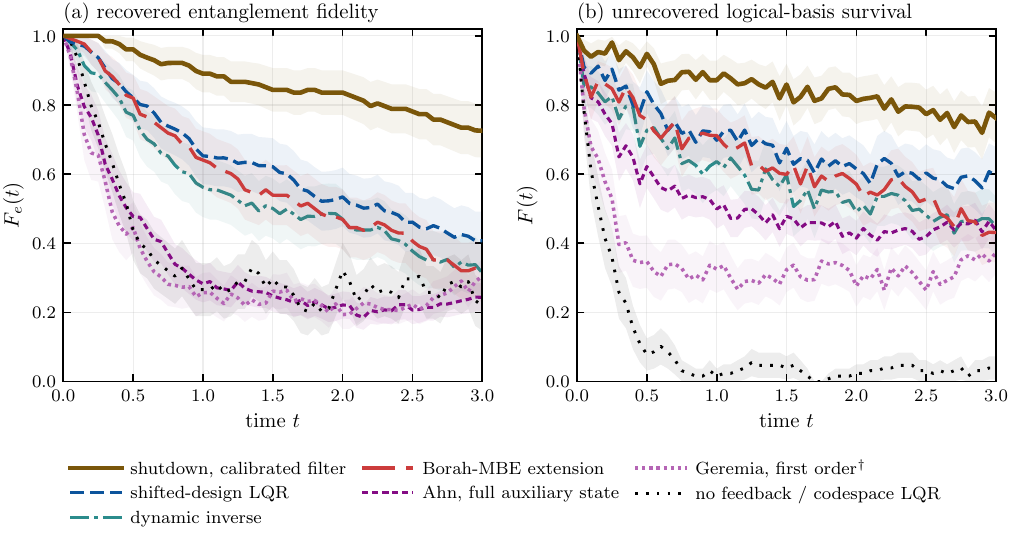}
\caption{Five-qubit benchmark. (a) Entanglement fidelity after ideal terminal recovery and decoding. (b) Unrecovered logical-basis survival. One representative per family: calibrated shutdown with the syndrome-only filter, shifted-design LQR without a floor, dynamic inversion, the Borah-MBE extension, Ahn, Geremia, and no feedback; the other arms are in Table~\ref{tab:unified-five}. Bands are pointwise 95\% intervals. Peak constraints are common; integrated expenditure is not matched. $\dagger$ Geremia's inferred joint probabilities leave $[0,1]$ in four test trajectories; none is discarded.}
\label{fig:unified-five}
\end{figure*}

\section{Scaling the pipeline: the five-qubit perfect code}
\label{sec:fivequbit}

No element of the loop (closure recursion, sector-wise bilinear models, gain synthesis, belief update, and the two interventions) depends on the bit-flip code. We now apply the same procedure to the five-qubit perfect code $[[5,1,3]]$~\cite{laflamme_perfect_1996,bennett_mixed_1996}, which corrects arbitrary single-qubit errors and therefore involves several noise species, a richer syndrome geometry, and a full set of Pauli actuators.

\subsection{Code, monitoring, actuation, and noise}
The code is stabilized by~\cite{gottesman_stabilizer_1997}
\begin{align}
g_1&=X_1Z_2Z_3X_4, & g_2&=X_2Z_3Z_4X_5, \nonumber\\
g_3&=X_1X_3Z_4Z_5, & g_4&=Z_1X_2X_4Z_5,
\end{align}
each monitored with strength $\kappa$ as in Eq.~\eqref{eq:sme}. The noise is single-qubit depolarizing, with fifteen jump operators $\{\sqrt{\gamma_5}\,P_i\}$, $P\in\{X,Y,Z\}$, $i=1,\dots,5$. Since every single-qubit Pauli error has a distinct syndrome, the actuation spans all fifteen channels,
\begin{equation}
H_c(t)=\frac{1}{2}\sum_{i=1}^{5}\sum_{P\in\{X,Y,Z\}} u_{i,P}(t)\, P_i,
\label{eq:5q-control}
\end{equation}
the sector estimator tracks 16 syndrome classes, and each nontrivial class activates the channel prescribed by lookup decoding.

We compare the two codes at equal physical error rate per qubit. Equal rate per channel would make each five-qubit qubit three times noisier, so we set
\begin{equation}
3\,\gamma_{5}=1\cdot\gamma_{3q}
\quad\Longrightarrow\quad
\gamma_{5}=\gamma_{3q}/3 =1/3,
\label{eq:matched-rate}
\end{equation}
which gives a register jump rate of 5 against 3 for the bit-flip code. With $T=3$ in both cases, the five-qubit register receives 15 expected error events against 9. The larger code must earn back this extra exposure. Matching the total register rate instead ($\gamma_5=0.2$) would make each five-qubit qubit $40\%$ cleaner and advantage the larger code. The decision interval is $h=2\times10^{-4}$ and register plus reference form a 64-amplitude vector.
\begin{table*}[t]
\caption{Frozen settings and held-out five-qubit endpoints at $T=3$ ($N=128$ per method). $\theta$ is the posterior-confidence threshold of threshold and pulse rules; a dash denotes the common continuous-controller selector. Amplitudes are the selected bound, not an increase of the common limit.}
\label{tab:unified-five}
\begin{ruledtabular}
\begin{tabular}{lrrrrll}
Method & $\kappa_{\rm det}$ & $\kappa_{\rm cor}$ & $u_{\max}^{\rm used}$ & $\theta$ & $F_e(3)$ & $F(3)$\\
Shutdown, calibrated & 400 & 0 & 400 & 0.995 & $0.7255\pm0.0395$ & $0.7621\pm0.0376$\\
Shutdown, tempered & 400 & 0 & 400 & 0.80 & $0.6852\pm0.0411$ & $0.7243\pm0.0393$\\
Floored LQR, fixed & 400 & 400 & 400 & -- & $0.3063\pm0.0408$ & $0.4172\pm0.0435$\\
Floored LQR, switched & 400 & 50 & 400 & -- & $0.4061\pm0.0436$ & $0.5106\pm0.0441$\\
Shifted-design LQR & 400 & 50 & 400 & -- & $0.4072\pm0.0429$ & $0.5963\pm0.0418$\\
Dynamic inverse & 100 & 100 & 400 & -- & $0.3145\pm0.0407$ & $0.4492\pm0.0434$\\
Ahn & 100 & 100 & 400 & -- & $0.2433\pm0.0205$ & $0.4376\pm0.0254$\\
Geremia first order$^{\dagger}$ & 200 & 200 & 100 & -- & $0.3049\pm0.0244$ & $0.3729\pm0.0312$\\
Borah-MBE extension & 400 & 400 & 400 & 0.50 & $0.3064\pm0.0408$ & $0.4312\pm0.0438$\\
Measurement-on pulse & 400 & 400 & 400 & 0.80 & $0.2500\pm0.0384$ & $0.3756\pm0.0426$\\
No feedback / codespace LQR & 400 & 400 & 400 & -- & $0.2188\pm0.0367$ & $0.0391\pm0.0172$\\
\end{tabular}
\end{ruledtabular}
\end{table*}
\subsection{Reduced filter and estimators}
\label{sec:native-five}
The recursion~\eqref{eq:general-basis-recursion} seeded with $\{g_1,\dots,g_4\}$ reproduces Theorem~\ref{thm:closure} with $n-k=4$: for each of the fifteen channels the closure has dimension $3\cdot2^{3}=24$, the sixteen stabilizer-group elements and the eight elements of $P\cdot\mathcal{S}^-$. Closing under all fifteen channels at once would yield $495$ operators, against $1023$ density-operator parameters.

The continuous LQR and inversion controllers use, for each recovery $P_q$, the 24-coordinate branch (23 independent moments), with precomputed operator maps, the positive split updates of Appendix~\ref{app:split}, and the sector projection~\eqref{eq:sector-projection} at axis changes; the last branch is kept while idle. They use the tempered selector with minimum nontrivial mass 0.10 and margin 0.05. Floored feedback uses recovery-adapted signs, $p_j^{\rm eval}=c_j\operatorname{sgn}^{+}(c_j\widehat p_j)\max(|\widehat p_j|,0.2)$. The shifted gain is designed at $s^\circ=0.98$, $p_j^\circ=c_j\sqrt{1-0.98^2}$ and evaluated with stabilizer reference 0.97; the codespace gain uses reference 1.

Calibrated shutdown uses the filter~\eqref{eq:calibrated-filter} with $C=16$, rate $\gamma_5$, and $p_a(0)=\delta_{a0}$. During shutdown only the noise prediction continues, and the permutation~\eqref{eq:completion} is applied at physical completion. The decision needs sixteen probabilities and no coherent partners. Dense register--reference checks validate the implemented split map and completion update (Appendix~\ref{app:unified-checks}). Residual measurement during a partial rotation violates Theorem~\ref{thm:shutdown}, so the syndrome-only update is then an approximation, tested in Sec.~\ref{sec:native-robustness}. The measurement-on pulse and the tempered shutdown keep the tempered posterior.

The Ahn and Geremia representations contain 1024 and 136 Pauli coordinates, including the identity, and Borah's auxiliary state has 1024. The published local-$Z$ detector has zero expectation in every five-qubit codespace and Pauli-error sector, since $P_C\sigma_qP_C=0$. We therefore replace it with a syndrome-confidence decision rule and call the result the \textit{Borah-MBE extension}: a full-state estimator with an adapted detector, distinct from the literal three-qubit protocol of Sec.~\ref{sec:three-unified}.

\paragraph*{Tuning.}
The training campaign has 65 candidates with 32 trajectories each, over measurement strengths 100, 200, and 400, amplitudes up to 400, confidence thresholds, and LQR and inversion parameters, plus an additional low-amplitude grid for Ahn and Geremia. For calibrated shutdown we first replace the filter at the earlier threshold 0.8, then select $\theta\in\{0.8,0.95,0.99,0.995,0.999\}$ on a separate 64-trajectory training ensemble with detection strength and drive fixed at 400, and freeze the choice before testing. This separates the filter replacement from the threshold retuning. Each test arm has 128 trajectories at $\delta t=10^{-5}$, and the refinement uses 64 independent trajectories at $5\times10^{-6}$.

Calibrated shutdown, at its selected threshold $\theta=0.995$, has the highest endpoints, $F_e(3)=0.7255\pm0.0395$ and $F(3)=0.7621\pm0.0376$ (Fig.~\ref{fig:unified-five}, Table~\ref{tab:unified-five}). The tempered shutdown gives $0.6852\pm0.0411$; the paired difference $0.0403$, interval $[-0.0264,0.1070]$, is unresolved. What changes is that good performance now comes with an estimator that is exact under ideal shutdown. Replacing the estimator while keeping threshold 0.8 gives only $F_e(3)=0.3672\pm0.0428$: exact syndrome information does not make a threshold tuned for another filter optimal, and training favors a higher confidence before committing to a blind interval.

Floored LQR with reduced measurement gives $F_e(3)=0.4061$ against $0.3063$ at fixed measurement, a paired gain of $0.0998$ that is positive in its ordinary interval but unresolved after adjustment (Sec.~\ref{subsec:switch-scope}). Shifted-design LQR gives $F_e(3)=0.4072\pm0.0429$ and $F(3)=0.5963\pm0.0418$ and matches floored LQR at the same schedule (paired difference $0.0010$, interval $[-0.0925,0.0946]$). The tempered measurement-on pulse reaches $F_e(3)=0.2500$, and switching the measurement off with the same filter and trigger gains $0.4352$, interval $[0.3295,0.5409]$; comparing calibrated shutdown with that pulse changes estimator and threshold as well.

The Borah-MBE extension gives $F_e(3)=0.3064\pm0.0408$ at strength 400 and drive 400. Calibrated shutdown exceeds it by $0.4192$, seven-contrast adjusted interval $[0.2842,0.5542]$, with mean measurement expenditure 4610.24 against 4800 and drive area 46.57 against 134.83. Ahn's adaptation gives $F_e(3)=0.2433$ despite a codespace population of 0.9008, which shows that returning to the codespace is not sufficient. Geremia's first-order adaptation gives $F_e(3)=0.3049$, but its inferred joint probabilities leave $[0,1]$ in four of 128 trajectories (maximum excursion 0.1333); this curve diagnoses our implementation of the approximation and does not show that the published approach is inferior.

At half the integration step, the independent 64-trajectory ensemble gives calibrated shutdown $F_e(3)=0.6875\pm0.0584$, and repeating the 128 main seeds gives $0.7471\pm0.0384$, a paired change of $0.0216$, interval $[-0.0454,0.0886]$. Full endpoints, the second metric, and resources are in Appendix~\ref{app:unified-checks}.

\begin{figure*}[tp]
\includegraphics[width=\textwidth]{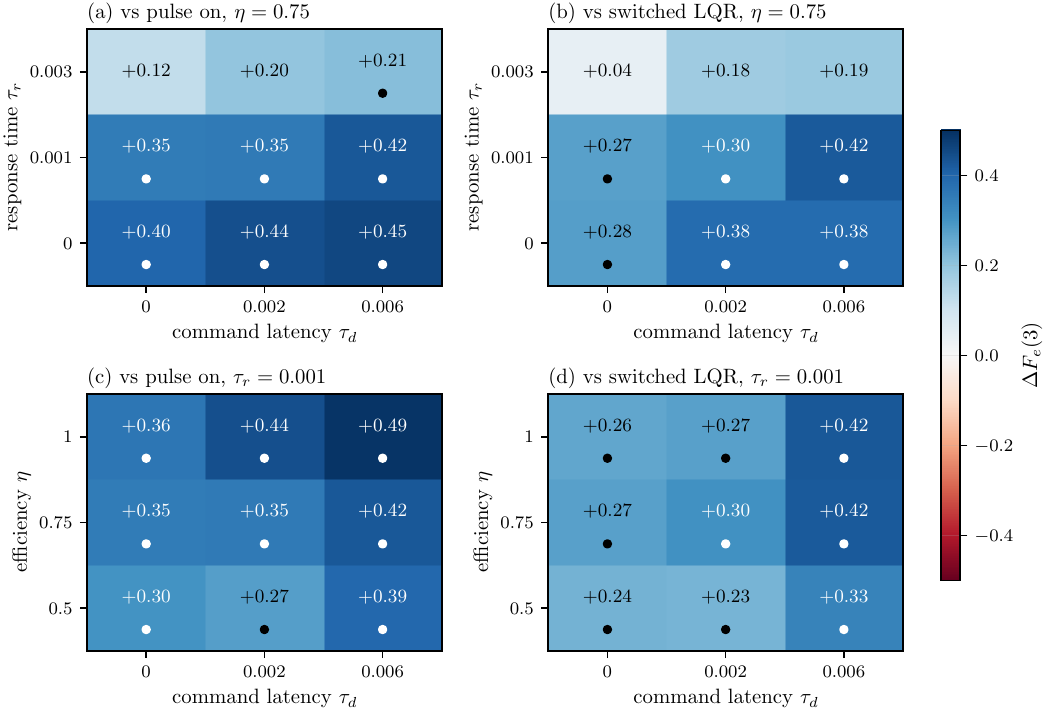}
\caption{Calibrated-shutdown operating cells. Entries give terminal $\Delta F_e(3)$ against the frozen comparator; a dot marks an adjusted lower 95\% bound above 0.02, with thirty paired contrasts in one Bonferroni family. (a,b) Latency versus response time at $\eta=0.75$. (c,d) Efficiency versus latency at $\tau_r=0.001$. Comparators are the tempered measurement-on pulse (a,c) and switched floored LQR (b,d). Cells are sampled, not interpolated; nonzero-response cells lie outside the exact theorem.}
\label{fig:native-robustness}
\end{figure*}

\section{Robustness to finite hardware response}
\label{sec:native-robustness}
We test calibrated shutdown at $\theta=0.995$ against two frozen comparators, the tempered measurement-on pulse and switched floored LQR, across fifteen hardware cells. All policies keep detection strength 400, drive cap 400, and correction targets 0, 400, and 50. Only the earlier tempered on/off pair isolates measurement scheduling at fixed estimator and trigger; these are controller--estimator comparisons.

A command queue delays both drive and measurement-target commands by $\tau_d$, a latency of the kind set by the feedback electronics~\cite{salathe_lowlatency_2018}. The estimator continues from current observations and applied controls, and the permutation is applied at physical completion. The measurement strength relaxes as
\begin{equation}
 \dot\kappa_{\rm act}=(\kappa_{\rm target}-\kappa_{\rm act})/\tau_r,
 \label{eq:measurement-response}
\end{equation}
starting at 400, a first-order stand-in for the finite ring-up and ring-down of a dispersive readout resonator~\cite{mcclure_rapid_2016}; $\tau_r=0$ is instantaneous switching. The filter knows the applied strength. At nonzero $\tau_r$ the residual measurement during recovery places the cell outside Theorem~\ref{thm:shutdown}. At $\eta=0.75$ we cross $\tau_d\in\{0,0.002,0.006\}$ with $\tau_r\in\{0,0.001,0.003\}$, and at $\tau_r=0.001$ the same delays with $\eta\in\{0.5,0.75,1\}$, giving fifteen cells with 128 trajectories per policy at $T=3$, $h=2\times10^{-4}$, and $\delta t=10^{-5}$.

\begin{table}[b]
\caption{Sampled cells favorable to calibrated shutdown against both frozen comparators. Unlisted cells are unresolved or unfavorable under the stated criterion.}
\label{tab:native-region}
\begin{ruledtabular}
\begin{tabular}{ccc}
$\eta$ & $\tau_r$ & Tested $\tau_d$\\
0.5 & 0.001 & 0, 0.002, 0.006\\
0.75 & 0 & 0, 0.002, 0.006\\
0.75 & 0.001 & 0, 0.002, 0.006\\
1 & 0.001 & 0, 0.002, 0.006\\
\end{tabular}
\end{ruledtabular}
\end{table}

We call a cell favorable when both adjusted lower bounds on $\Delta F_e(3)$ exceed 0.02, with $z=\Phi^{-1}(1-0.05/60)$. The criterion holds at 12 of the fifteen cells (Fig.~\ref{fig:native-robustness}, Table~\ref{tab:native-region}), and no trajectory fails. At $\tau_r=0.003$ no cell meets it, consistent with residual monitoring impairing coherent recovery. A half-step check with 64 independent trajectories per arm gives calibrated shutdown $F_e(3)=0.7500\pm0.0546$, $0.6562\pm0.0598$, and $0.3594\pm0.0605$ at $(\tau_d,\tau_r)=(0,0)$, $(0.002,0.001)$, and $(0.006,0.003)$, all at $\eta=0.75$; both refined gain intervals are positive at the first two points. The best-supported nonideal operating point is therefore $\eta=0.75$, $\kappa_{\rm det}=u_{\max}=400$, $\tau_d=0.002$, and $\tau_r=0.001$. The grid also favors shutdown at $\eta=0.5$, a slice without refinement. These results identify tested operating points; they do not determine a continuous boundary or a critical efficiency.

\section{Discussion and outlook}
\label{sec:discussion}
Continuous quantum error correction offers lower thresholds and reduced hardware overhead compared to its digital counterpart, but these advantages have so far been offset by the difficulty of estimating and controlling the underlying stochastic model. In this work, we have shown that this difficulty is largely one of formulation. By recasting continuous QEC as a hierarchy of classical control-theoretic objects (an exact bilinear reduced model, an exact conditional filter, and estimation and feedback laws built on top of them), we replaced an ad hoc modeling problem with a sequence of well-posed steps, each of which can be addressed with established tools from classical estimation and control.

The sector model exposes two bottlenecks: codespace-linearized LQR can lack an ignition command, and strong monitoring damps the partners needed for coherent recovery. A gain designed at a shifted operating point, or equivalently a floor, addresses the first; reduced measurement during correction and committed shutdown address the second. The benefit of reduced measurement for continuous LQR grows with the number of flipped monitored channels: unresolved for the two-generator three-qubit sectors, resolved when all three parities are monitored, and favored in point estimate for the five-qubit code. Committed shutdown is the strongest law in both codes.

The two theorems have different scopes. I. Fixed-channel closure includes the partners needed during coherent motion. II. Under measurement-off recovery and covariant noise, the shutdown-boundary identity allows exact syndrome tracking between pulses without those partners, and the calibrated-filter experiments test that architecture directly. Its confidence threshold must be selected afresh, because exact estimation does not by itself optimize recovery timing. In three qubits, calibrated shutdown exceeds the published Borah protocol under identical three-parity monitoring; with only the two generator currents its difference from the three-current protocol is unresolved. An earlier three-qubit study measured a 3.26-fold reduction in online computation time for the nine-coordinate pulse filter relative to full Borah (Appendix~\ref{app:three-borah}). We do not transfer this ratio to the five-qubit filter, where the Borah detector is adapted; that claim would need a matched-record, complete-loop timing experiment.

\paragraph*{The redundant parity: information versus measurement cost.}
Monitoring $Z_1Z_3$ in addition to the two generators raises every leading three-qubit method (Table~\ref{tab:parity-cost}): by $0.041$, interval $[0.019,0.063]$, for calibrated shutdown; $0.098$, interval $[0.058,0.138]$, for shifted-design LQR; and $0.104$, interval $[0.068,0.139]$, for the Borah rule, whose two-generator version is tuned separately ($\kappa=\lambda=200$ instead of 400). The third parity carries no new syndrome information but increases the overall speed of the estimator. Two classes whose record means differ in $d$ channels are separated by log-likelihood evidence accumulating at rate $8\eta\kappa d$~\cite{wald_sequential_1945}, since each differing channel shifts the mean by $4\sqrt{\eta\kappa}$. With the generators alone, the codespace differs from $X_1$ or $X_3$, and $X_2$ from $X_1$ or $X_3$, in a single channel, so $d_{\min}=1$; in contrast, with all three parities every pair differs in two channels, and errors are identified about twice as fast.

In the bit-flip code, this speed determines whether a physical error becomes a logical one. An error is fatal if a second error hits another qubit, at rate $2\gamma$, before the first is identified. Reaching $\theta=0.99$ against the small prior odds of a fresh error takes of order ten nats of evidence, which at $\eta=0.75$ and $\kappa=400$ means delays of about $4\times10^{-3}$ for $d=1$ and $2\times10^{-3}$ for $d=2$. Over the $3\gamma T=9$ expected errors, the fatal fractions predict losses of roughly 8\% and 4\%; calibrated shutdown loses 9.4\% and 5.3\%. This order-of-magnitude estimate is not a fit, but it accounts for most of the difference.

\begin{table}[t]
\caption{Effect of the redundant parity $Z_1Z_3$ in the three-qubit comparison. $d$ counts the monitored channels whose record means differ between two syndrome classes; the evidence separating them accumulates at rate $8\eta\kappa d$. Endpoints are held-out $F_e(3)$ with paired error histories; $B_\kappa$ and $A_u$ refer to calibrated shutdown.}
\label{tab:parity-cost}
\begin{ruledtabular}
\begin{tabular}{lcc}
 & Two generators & Three parities\\
$d(0,X_1)=d(0,X_3)$ & 1 & 2\\
$d(0,X_2)$ & 2 & 2\\
$d(X_1,X_2)=d(X_2,X_3)$ & 1 & 2\\
$d(X_1,X_3)$ & 2 & 2\\
Calibrated shutdown & $0.9062\pm0.0129$ & $0.9473\pm0.0099$\\
Shifted-design LQR & $0.7657\pm0.0186$ & $0.8638\pm0.0151$\\
Borah local-$Z$ rule & $0.8164\pm0.0171$ & $0.9199\pm0.0120$\\
$B_\kappa$, calibrated shutdown & 2289 & 3432\\
$A_u$, calibrated shutdown & 27.5 & 27.8\\
\end{tabular}
\end{ruledtabular}
\end{table}

To continue improving the framework, future work should extend the boundary analysis to residual measurement, compare a broader range of implementations and tuning grids, and evaluate longer trajectories to assess logical lifetime and fault tolerance. The model should also incorporate pulse-calibration errors, non-Markovian noise, measurement-induced dephasing of realistic parity meters~\cite{halaski_obstacles_2026}, and hardware-specific timing.  An open direction is to include measurement-independent drives to improve robustness and fixed-point fidelities \cite{greenfield2024stabilizing}. Within these current limits, the analytic and numerical results strongly support scheduling the observation around bounded coherent recovery and accounting for the completed control in the reduced estimator.

\section{Acknowledgement}
We thank Francesco Ticozzi, Tommaso Grigoletto, Jos\'e Da Costa Jesus and Dimitrios Georgiadis for valuable discussions and insightful feedback. This work was supported by the European Union’s Horizon Europe programme through the QCFD project (Grant No. 101080085, HORIZON-CL4-2021-DIGITAL-EMERGING-02-10),  OpenSuperQPlus100 (Grant No. 101113946, HORIZON-CL4-2022-QUANTUM-01-SGA), and PASQuanS2.1 (Grant No. 101113690, HORIZON-CL4-2022-QUANTUM-02-SGA), the ML4Q2 Cluster of Excellence (EXC 2004/2 – 390534769), and the German Federal Ministry of Education and Research (BMBF) through the QSolid project (Grant No. 13N16149) and  MUNIQC-Atoms (Grant No. 13N16073).

\clearpage

\appendix

\section{Proof of Theorem~\ref{thm:closure}}
\label{app:closure-proof}

\begin{proof}
With $L_\mu^2=g_j^2=P^2=I$, the three generating operations act on an observable $O$ as follows.

\begin{center}
\renewcommand{\arraystretch}{1.3}
\begin{tabular}{@{}llll@{}}
\hline\hline
Operation & Acts as & Non-zero iff & New element \\
\hline
Dissipative drift  & $L_\mu OL_\mu-O$ & $\{L_\mu,O\}=0$ & none ($\propto O$) \\
Meas. update & $g_jO+Og_j$      & $[g_j,O]=0$     & $g_jO$ \\
Control            & $i[P,O]$         & $\{P,O\}=0$     & $PO$ \\
\hline\hline
\end{tabular}
\end{center}

\noindent The noise never enlarges the basis, so the closure is independent of the noise model. The recursion is the orbit of the seeds under left multiplication by $\langle\mathcal{S},P\rangle$, with each multiplication allowed only under its commutation condition. We show that $\mathcal{B}$ of Eq.~\eqref{eq:closure-set} is invariant under both multiplications and is generated by the seeds.

\textit{Invariance, first piece.} Let $O=s\in\mathcal{S} \setminus I$. Since $\mathcal{S}$ is abelian, the measurement update always fires and returns $g_js\in\mathcal{S}$; the trivial result $g_js=I$ is discarded. The control fires only if $s\in\mathcal{S}^-$ and then returns $Ps\in P\cdot\mathcal{S}^-$.

\textit{Invariance, second piece.} Let $O=Ps$ with $s\in\mathcal{S}^-$. Since $[Ps, g_j] = P[s,g_j] + [P,g_j]s = [P,g_j]s$, the measurement update fires exactly when $[P,g_j] = 0$, that is $g_j\in\mathcal{S}^+$. It then gives $g_jPs + Psg_j = 2Psg_j = 2Ps'$ with $s' = sg_j \in \mathcal{S}^-\!\cdot\mathcal{S}^+ = \mathcal{S}^-$, again in $P\cdot\mathcal{S}^-$. The control fires because $s \in \mathcal{S}^-$, and $[P,Ps] = 2s\in \mathcal{S}^-$ lies in the first piece.

\textit{Generation.} Repeated measurement updates of the seeds produce every product of generators and hence all of $\mathcal{S}\setminus I$; no intermediate product equals $I$ because the generators are independent. Applying the control to every $s\in\mathcal{S}^-$ then produces all of $P\cdot\mathcal{S}^-$.

\textit{Cardinality.} $|\mathcal{S}\setminus I| = 2^{n-k}-1$, and $s\mapsto Ps$ is injective because $P^2=I$, so $|P\cdot\mathcal{S}^-|=|\mathcal{S}^-|$. Fix $t\in \mathcal{S}^-$, which exists by hypothesis, and define $\psi : \mathcal{S}^+ \rightarrow \mathcal{S}^-$, $\psi(s) = ts$. It is well defined, since $P(ts) = -t(Ps) = -(ts)P$; injective, since $t^2=I$; and surjective, since for $m \in \mathcal{S}^-$ the element $tm$ commutes with $P$ and $\psi(tm)=m$. The sets $\mathcal{S}^\pm$ partition $\mathcal{S}$, because any two Paulis commute or anticommute and $sP=-sP$ is impossible, so $|\mathcal{S}^-| = \tfrac{1}{2}|\mathcal{S}| = 2^{\,n-k-1}$.

Finally, the two pieces are disjoint. If $Ps\in\mathcal{S}$ for some $s\in\mathcal{S}^-$, then $P=(Ps)s\in\mathcal{S}$ would commute with all of $\mathcal{S}$, against the hypothesis. The cardinalities add to $(2^{n-k}-1)+2^{n-k-1} = 3\cdot 2^{n-k-1}-1$.
\end{proof}

\section{Nine-coordinate union filter for the three-qubit code}
\label{app:9dfilter}

The fixed-channel span of Theorem~\ref{thm:closure} depends on the active recovery and does not contain the logical operators. The earlier three-qubit timing study (Appendix~\ref{app:three-borah}) instead propagates one sector-independent model on
\begin{equation}
x=\begin{bmatrix}
s_{12} & s_{23} & s_{13} & p_1 & p_2 & p_3 & p_4 & q_1 & q_2
\end{bmatrix}^{\top},
\label{eq:9d-coordinates}
\end{equation}
with $s_{12}=\ev{Z_1Z_2}$, $s_{23}=\ev{Z_2Z_3}$, $s_{13}=\ev{Z_1Z_3}$, and the six partners
\begin{align}
p_1&=\ev{Y_1Z_2}, & p_2&=\ev{Z_1Y_2}, & p_3&=\ev{Y_2Z_3}, \nonumber\\
p_4&=\ev{Z_2Y_3}, & q_1&=\ev{Y_1Z_3}, & q_2&=\ev{Z_1Y_3},
\label{eq:9d-partners}
\end{align}
the union of $P\cdot\mathcal{S}^-$ over $P\in\{X_1,X_2,X_3\}$; $q_1$ and $q_2$ are the partners of $s_{13}$ under $X_1$ and $X_3$. One set of coordinates then serves all four sectors, against 63 real parameters for the density operator. With the innovations $dW_{12}=dY_{12}-2\sqrt{\eta\kappa}\,s_{12}dt$ and $dW_{23}=dY_{23}-2\sqrt{\eta\kappa}\,s_{23}dt$, noise and measurement give the diagonal damping of Eq.~\eqref{eq:drift-rule},
\begin{align}
\dot{s}_{12}&=-4\gamma s_{12}, &
\dot{s}_{23}&=-4\gamma s_{23}, \nonumber\\
\dot{s}_{13}&=-4\gamma s_{13}, &
\dot{p}_1&=-(4\gamma+2\kappa) p_1, \nonumber\\
\dot{p}_2&=-(4\gamma+4\kappa) p_2, &
\dot{p}_3&=-(4\gamma+4\kappa) p_3, \nonumber\\
\dot{p}_4&=-(4\gamma+2\kappa) p_4, &
\dot{q}_1&=-(4\gamma+2\kappa) q_1, \nonumber\\
\dot{q}_2&=-(4\gamma+2\kappa) q_2, &&
\label{eq:9d-damping}
\end{align}
and the coherent terms are
\begin{align}
\dot{s}_{12}&=u_1p_1+u_2p_2, & \dot{p}_1&=-u_1s_{12}, \nonumber\\
\dot{p}_2&=-u_2s_{12}, & \dot{s}_{23}&=u_2p_3+u_3p_4, \nonumber\\
\dot{p}_3&=-u_2s_{23}, & \dot{p}_4&=-u_3s_{23}, \nonumber\\
\dot{s}_{13}&=u_1q_1+u_3q_2, & \dot{q}_1&=-u_1s_{13}, \nonumber\\
\dot{q}_2&=-u_3s_{13}. &&
\label{eq:9d-control}
\end{align}
The stochastic part of the filter is
\begin{align}
ds_{12} &= 2\sqrt{\eta\kappa}(1-s_{12}^2)dW_{12} \nonumber\\
&\quad +2\sqrt{\eta\kappa}(s_{13}-s_{12}s_{23})dW_{23}, \nonumber\\
ds_{23} &= 2\sqrt{\eta\kappa}(1-s_{23}^2)dW_{23} \nonumber\\
&\quad +2\sqrt{\eta\kappa}(s_{13}-s_{12}s_{23})dW_{12}, \nonumber\\
ds_{13} &= 2\sqrt{\eta\kappa}(s_{23}-s_{12}s_{13})dW_{12} \nonumber\\
&\quad +2\sqrt{\eta\kappa}(s_{12}-s_{23}s_{13})dW_{23}, \nonumber\\
dp_1 &= -2\sqrt{\eta\kappa}s_{12}p_1dW_{12} \nonumber\\
&\quad +2\sqrt{\eta\kappa}(q_1-s_{23}p_1)dW_{23}, \nonumber\\
dp_2 &= -2\sqrt{\eta\kappa}s_{12}p_2dW_{12} \nonumber\\
&\quad -2\sqrt{\eta\kappa}s_{23}p_2dW_{23}, \nonumber\\
dp_3 &= -2\sqrt{\eta\kappa}s_{12}p_3dW_{12} \nonumber\\
&\quad -2\sqrt{\eta\kappa}s_{23}p_3dW_{23}, \nonumber\\
dp_4 &= 2\sqrt{\eta\kappa}(q_2-s_{12}p_4)dW_{12} \nonumber\\
&\quad -2\sqrt{\eta\kappa}s_{23}p_4dW_{23}, \nonumber\\
dq_1 &= -2\sqrt{\eta\kappa}s_{12}q_1dW_{12} \nonumber\\
&\quad +2\sqrt{\eta\kappa}(p_1-s_{23}q_1)dW_{23}, \nonumber\\
dq_2 &= 2\sqrt{\eta\kappa}(p_4-s_{12}q_2)dW_{12} \nonumber\\
&\quad -2\sqrt{\eta\kappa}s_{23}q_2dW_{23},
\label{eq:closed-filter}
\end{align}
to which the deterministic terms of Eqs.~\eqref{eq:9d-damping} and~\eqref{eq:9d-control} are added.

Every innovation coefficient is a polynomial in the nine coordinates, so the stochastic part is exact. The coherent part does not close globally: a recovery $X_i$ maps the partners of the other recoveries outside the span, for instance $i[X_1,Z_1Y_2]=2Y_1Y_2$, and those images are truncated. The active sector is unaffected. When $X_1$ drives, the coordinates it takes outside the span are $p_2$ and $q_2$, which enter the active equations only through $u_2p_2$ and $u_3q_2$, both zero; the block $\{s_{12},s_{23},s_{13},p_1,q_1\}$, the Theorem~\ref{thm:closure} set for $P=X_1$, evolves without approximation, and likewise for $X_2$ and $X_3$. The truncation costs only the freshness of off-sector partners, which are stale if the estimator switches mid-recovery. These are the most strongly damped coordinates, at $4\gamma+2\kappa w$, and the escaped operators $Y_iY_j$ decay at $4\gamma+2\kappa$, so at $\kappa\gtrsim100\gamma$ the error is suppressed within a few macro-steps, against recovery transients two orders of magnitude longer.

\section{Sign convention, controllability, and polar form of the sector model}
\label{app:polar}

\paragraph*{Sign convention.}
A filter carried across all sectors names its observables once, independently of the active recovery. Write $P = \sigma_a$ on the corrected qubit $q$ and $g_j  = \sigma_{b_j}^{(q)} R_j$, with $R_j$ supported away from $q$ and commuting with $P$; anticommutation forces $b_j \neq a$. Then
\begin{equation}
  \tilde g_j = i\,\sigma_a \sigma_{b_j} \otimes R_j
             = -\varepsilon_{a b_j c}\,\tilde P_j ,
  \qquad
  \tilde P_j := \sigma_c^{(q)} R_j ,
  \label{eq:partner-sign}
\end{equation}
using $\sigma_a \sigma_b = i\,\varepsilon_{abc}\sigma_c$ for $a \neq b$, so $\tilde g_j = c_j \tilde P_j$ with
\begin{equation}
  c_j = -\varepsilon_{a b_j c} = \pm 1 .
  \label{eq:cj}
\end{equation}
Both operators are Hermitian and square to the identity, so no other phase is possible. The sign is fixed by which of the two Pauli types anticommuting with $\sigma_a$ the generator carries at $q$.

The remaining properties concern the unconditional model on which the gains are synthesized. For a single plane $(s,p)$ with $j\in F$, write $\Delta := a_p - a_s$ with the rates of Eq.~\eqref{eq:sector-rates}, so that
\begin{equation}
\dot s = -a_s\,s + u\,p ,
\qquad
\dot p = -a_p\,p - u\,s .
\label{eq:polar-plane}
\end{equation}
For depolarizing noise and for the bit-flip code, $\Delta = 2\kappa w$.

\paragraph*{Controllability of the surrogate.}
The determinant of Eq.~\eqref{eq:ctrb-det} is $b_1b_2(a_s-a_p)$, so the pair is controllable only if the operating point is shifted and $\Delta\neq0$, which holds through the measurement term alone. For $w>1$ the $2w$-dimensional pair has a controllability matrix of rank two, since the drift carries $a_s$ and $a_p$ each with multiplicity $w$. The deficiency is inert. A syndrome error leaves every flipped generator at $s_j=-1$ with a zero partner, so all planes start in the same state, and the synchronized set $\{(s,c_1p,\dots,s,c_wp)\}$ is invariant under drift and control, with dynamics given exactly by Eq.~\eqref{eq:polar-plane}. This is why sectors with $w$ up to 4 are corrected in Sec.~\ref{sec:fivequbit}.

\paragraph*{Stabilizability.}
The drift is Hurwitz for every $w$, so the pair is stabilizable for any input direction, including $b_{\mathrm{lin}}=0$, and the test is passed by a controller that does nothing. The uncontrolled model converges to the origin, the fully depolarized state, so stabilization is not the objective. The only use of the property is that every uncontrollable mode is open-loop stable, so Eq.~\eqref{eq:Ricatti} has a stabilizing solution that gives the undriven directions zero gain, as used in Eq.~\eqref{eq:codespace-gain}.

\paragraph*{Gain structure for $w>1$.}
At the codespace point $b_{\mathrm{lin}}$ has a nonzero entry on every partner, so $K$ picks up the off-diagonal partner blocks of the Riccati solution. The stabilizer gains still vanish, and since all planes share their rates the partner gains are equal in magnitude with signs $c_j$, giving $u=k_p\sum_j c_j\hat p_j$ with
\begin{equation}
k_p=\frac{s^{\circ}}{R}\Bigl(P_{pp}+(w-1)\,P_{p_lp_j}\Bigr),
\qquad l\neq j,
\label{eq:kp-general}
\end{equation}
where $P_{pp}$ and $P_{p_lp_j}$ are the common diagonal and off-diagonal partner entries. The off-diagonal term is negative, so $k_p$ decreases with multiplicity: at $\kappa=200\gamma$ and the design values of Table~\ref{tab:design-values}, $k_p/\gamma = 1868$, $1229$ and $787$ for $w=1$, $2$ and $4$.

\paragraph*{Polar form.}
With $s = r\cos\theta$ and $p = -r\sin\theta$, the sign chosen so that $\dot\theta = u$ when $a_s=a_p=0$ (the normalization enforced by the factor $\tfrac12$ in Eq.~\eqref{eq:control-hamiltonian}), $\theta=0$ is the codespace and $\theta=\pi$ the post-error point. Substituting Eq.~\eqref{eq:polar-plane}, the control terms cancel in $r\dot r = s\dot s + p\dot p$ and
\begin{align}
\dot r &= -r\bigl(a_s + \Delta\sin^2\theta\bigr) ,
\label{eq:rdot}\\
\dot\theta &= u - \tfrac{1}{2}\Delta\,\sin 2\theta .
\label{eq:thetadot}
\end{align}
The control contributes $u$ by construction. The second term exists only because the damping is anisotropic: the drift contracts the partner faster than the stabilizer and pulls the vector toward the slow axis. It vanishes for $\Delta=0$ and at the eigendirections $\sin2\theta=0$, and near $\theta=0$ it gives $\dot\theta\simeq u-\Delta\theta$.

\section{Sector-wise regularized dynamic inversion}
\label{subsec:sector_dynamic_inversion}

In one sector, $\dot{x} = A x + g(x) u$ with $g(x) := Bx$, so at fixed $x$ the scalar input moves the state only along $g(x)$. The derivative of the raw current is not a regular signal, so the controller uses a smoothed estimate and a filtered slope,
\begin{align}
\bar{x}_{k+1} &= (1-\alpha_x)\bar{x}_k + \alpha_x \hat{x}_k,
\label{eq:state_smoothing}\\
d_{k+1} &= (1-\alpha_d)d_k + \alpha_d \frac{\bar{x}_{k+1}-\bar{x}_k}{h},
\label{eq:derivative_smoothing}
\end{align}
with $d_k$ an estimate of $\dot{\bar{x}}$. It prescribes the closed-loop velocity
\begin{equation}
v_{\mathrm{des}}
:=
- K_p \bigl(x-x_{\mathrm{ref}}\bigr)
- K_d \dot{\bar{x}},
\label{eq:v_des_def}
\end{equation}
with positive semidefinite $K_p$ and $K_d$. Matching $Ax + g(x) u = v_{\mathrm{des}}$ requires $g(x)u = r(x) := v_{\mathrm{des}} - A x$, which a scalar $u$ can satisfy only if $r$ is collinear with $g$. The least-squares solution projects $r$ onto $g$, and Tikhonov regularization with $\lambda>0$ avoids the singularity at small $\norm{g}$:
\begin{equation}
u_{\mathrm{reg}}
=
\frac{g(x)^\top\left[- K_p (x-x_{\mathrm{ref}})- K_d \dot{\bar{x}}- A x\right]}{g(x)^\top g(x)+\lambda}.
\label{eq:ureg_full}
\end{equation}
The applied control is $u=\mathrm{sat}_{[-u_{\max},u_{\max}]}(u_{\mathrm{reg}})$ on the channel of the active sector, with the others set to zero; no correction is applied when the selector reports the codespace. At $\hat{p}=0$ the desired stabilizer restoration is orthogonal to $g(x)=(\hat p, -\hat s)^\top$, and the projected control vanishes (Sec.~\ref{sec:deadlock}). The benchmarks therefore use the operating-point variant, in which the direction is frozen at the shifted point~\eqref{eq:shifted-op}, $g_{\mathrm{eff}}=B\,x^{\circ}$ with $x^{\circ}=(s^{\circ},p_{\mathrm{op}})$, so the projection never degenerates.

\section{Shutdown proof, a noise bound, and numerical checks}
\label{app:shutdown-proof}
\begin{proof}[Proof of Theorem~\ref{thm:shutdown}]
Throughout, $r=n-k$, the syndrome of class $a\in\mathbb F_2^r$ has signs $s_{ja}=(-1)^{a_j}$, and
\begin{equation}
 \Pi_a=\prod_{j=1}^{r}\frac{I+s_{ja}\,g_j}{2},
 \qquad g_j\Pi_a=s_{ja}\Pi_a,
 \qquad \sum_a\Pi_a=I .
 \label{eq:syndrome-projectors}
\end{equation}
For a Pauli $E$, $s(E)\in\mathbb F_2^r$ has $s(E)_j=1$ exactly when $\{E,g_j\}=0$. The Pauli noise is $\mathcal L_N=\sum_\mu\gamma_\mu(\operatorname{Ad}_{E_\mu}-\mathrm{id})$ with Hermitian Paulis $E_\mu$, and $\operatorname{Ad}_E(\rho)=E\rho E$. An untouched reference $R$ only tensors every map with $\mathrm{id}_R$ and every projector with $I_R$; since $\operatorname{Tr}[(\Pi_a\otimes I_R)(\mathcal E\otimes\mathrm{id}_R)(\rho)]=\operatorname{Tr}[\Pi_a\,\mathcal E(\operatorname{Tr}_R\rho)]$, we suppress $R$ below.

\textit{Lemma.} (a) For any Pauli $E$, $E\Pi_aE=\Pi_{a\oplus s(E)}$. (b) For any state $\rho$ with populations $p_a=\operatorname{Tr}(\Pi_a\rho)$ and any $t\ge0$, $\operatorname{Tr}[\Pi_a e^{t\mathcal L_N}(\rho)]=(e^{tQ}p)_a$, with $Q$ of Eq.~\eqref{eq:syndrome-generator}.
Part (a) follows from $Eg_jE=(-1)^{s(E)_j}g_j$, which flips exactly the signs $s_{ja}$ with $s(E)_j=1$. For (b), let $p_a(t)=\operatorname{Tr}[\Pi_a e^{t\mathcal L_N}(\rho)]$. By (a), $\mathcal L_N^\dagger(\Pi_a)=\sum_\mu\gamma_\mu(\Pi_{a\oplus s(E_\mu)}-\Pi_a)$, so
$\dot p_a=\operatorname{Tr}[\mathcal L_N^\dagger(\Pi_a)\,e^{t\mathcal L_N}(\rho)]=(Qp(t))_a$. This linear system has the unique solution $p(t)=e^{tQ}p(0)$.

\textit{Recovery interval.} Shift time so that the pulse starts at $0$ and write $\theta(t)=\int_0^t u(t')\,dt'$. With the measurement off, the unobserved state obeys $\dot\rho=\mathcal G(t)\rho$ with $\mathcal G(t)=u(t)\mathcal H_P+\mathcal L_N$. Let $\mathcal F(t)=e^{t\mathcal L_N}e^{\theta(t)\mathcal H_P}$. Because $[\mathcal L_N,\mathcal H_P]=0$, $\mathcal H_P$ commutes with $e^{t\mathcal L_N}$, and
\begin{equation}
 \dot{\mathcal F}(t)=\mathcal L_N\mathcal F(t)+u(t)\,e^{t\mathcal L_N}\mathcal H_P\,e^{\theta(t)\mathcal H_P}=\mathcal G(t)\mathcal F(t),
 \qquad \mathcal F(0)=\mathrm{id}.
\end{equation}
For bounded measurable $u$, this linear equation has a unique absolutely continuous solution, so the propagator over the pulse is $\mathcal F(\tau)=e^{\tau\mathcal L_N}e^{\pi\mathcal H_P}$. Since $P^2=I$, $e^{-i\pi P/2}=-iP$ and $e^{\pi\mathcal H_P}(\rho)=(-iP)\rho(iP)=\operatorname{Ad}_P(\rho)$. Hence $\mathcal E_{\rm off}=\operatorname{Ad}_P\circ e^{\tau\mathcal L_N}$, the two factors commuting by assumption (iii). Applying Lemma (a) to $\operatorname{Ad}_P$ and then Lemma (b),
\begin{equation}
 p_a(t_1)=\operatorname{Tr}\bigl[\Pi_{a\oplus s(P)}\,e^{\tau\mathcal L_N}(\rho_c(t_0))\bigr]=\bigl(M_Pe^{\tau Q}p(t_0)\bigr)_a ,
\end{equation}
which is Eq.~\eqref{eq:shutdown-exact}. The populations after the pulse depend only on the populations before it, whatever coherences $\rho_c(t_0)$ contains.

\textit{Detection interval.} With zero Hamiltonian, the conditional state obeys
\begin{equation}
\begin{split}
 d\rho_c={}&\Bigl[\mathcal L_N(\rho_c)+\sum_j\kappa_j\mathcal D[g_j]\rho_c\Bigr]dt\\
 &+\sum_j\sqrt{\eta_j\kappa_j}\,\mathcal M[g_j]\rho_c\,dW_j,
\end{split}
\end{equation}
with innovations $dW_j=dY_j-2\sqrt{\eta_j\kappa_j}\,\langle g_j\rangle\,dt$. We take $\operatorname{Tr}(\Pi_a\,\cdot)$ term by term. The noise term gives $(Qp)_a\,dt$ by Lemma (b). The dephasing term vanishes, because $g_j\Pi_ag_j=s_{ja}^2\Pi_a=\Pi_a$ gives $\mathcal D^\dagger[g_j](\Pi_a)=0$. For the innovation term, Eq.~\eqref{eq:syndrome-projectors} gives $\operatorname{Tr}[\Pi_a(g_j\rho_c+\rho_cg_j)]=2s_{ja}p_a$ and $\langle g_j\rangle=\sum_bs_{jb}p_b=\bar s_j$. Together,
\begin{equation}
 dp_a=(Qp)_a\,dt+2p_a\sum_j\sqrt{\eta_j\kappa_j}\,(s_{ja}-\bar s_j)\,dW_j ,
\end{equation}
which is Eq.~\eqref{eq:exact-syndrome-filter}. Since $\langle g_j\rangle=\bar s_j$ is itself a function of $p$, the innovations, and with them the whole equation, are determined by the record $dY$ and $p$ alone. The coefficients are smooth and the probability simplex is invariant, so the equation has a unique strong solution given $p$ at the start of the interval. No assumption on the coherences of $\rho_c$ was used.

\textit{Adaptive chaining.} Let $\mathcal Y_t$ be the record up to time $t$. The $m$-th pulse starts at a time $t_0^{(m)}$ decided from the record, that is, a stopping time with respect to $(\mathcal Y_t)$, and its axis $P^{(m)}$ and profile $u^{(m)}$ are functions of $\mathcal Y_{t_0^{(m)}}$. During the pulse the measurement is off, so no record arrives and nothing new is learned. Conditioned on $\mathcal Y_{t_0^{(m)}}$, the pulse is therefore the fixed channel $\mathcal E_{\rm off}^{(m)}$ applied to $\rho_c(t_0^{(m)})$, and the conditional state at completion is $\rho_c(t_1^{(m)})=\mathcal E_{\rm off}^{(m)}(\rho_c(t_0^{(m)}))$. Each pulse lasts at least $\pi/u_{\max}$, so at most $\lceil u_{\max}T/\pi\rceil$ pulses fit in $[0,T]$. Induction over them, starting from the known initial populations, proves the claim. The detection result gives the exact populations up to $t_0^{(1)}$. The recovery result carries them exactly to $t_1^{(1)}$. Those are the initial condition of the next detection interval, to which the detection result applies again, and so on.

The equality concerns the conditional state, which averages over the unobserved noise jumps during each pulse. In a single realization with a jump during a pulse, the physical state differs from that average, and the theorem makes no statement about it.
\end{proof}

\begin{proposition}[Bound without noise covariance]
\label{prop:noise-bound}
Keep the ideal shutdown and a committed pulse $H_c(t)=u(t)P/2$ on $[t_0,t_0+\tau]$, with $u$ bounded and measurable and $\int_{t_0}^{t_0+\tau}u\,dt=\pi$, but allow arbitrary time-independent Pauli noise
\begin{equation}
 \mathcal L_N=\sum_\mu\gamma_\mu\bigl(\operatorname{Ad}_{E_\mu}-\mathrm{id}\bigr),
\end{equation}
with Hermitian Pauli operators $E_\mu$, rates $\gamma_\mu\ge0$, and $\operatorname{Ad}_E(\rho)=E\rho E$. Let $\Gamma_\perp=\sum_{\mu:\{E_\mu,P\}=0}\gamma_\mu$. If $\mathcal E_{\rm true}$ is the unobserved pulse channel and $\mathcal E_{\rm bdry}=\operatorname{Ad}_P\circ e^{\tau\mathcal L_N}$ its boundary approximation, then
\begin{equation}
 \tfrac12\|\mathcal E_{\rm true}-\mathcal E_{\rm bdry}\|_\diamond
 \le 1-e^{-\Gamma_\perp\tau}.
 \label{eq:shutdown-bound}
\end{equation}
Consequently, for every reference system $R$, every state $\rho$ of register and reference, every completely positive trace-preserving map $\mathcal F$ applied afterwards, and every operator $0\le\Pi\le I$,
\begin{equation}
\begin{split}
 \bigl|\operatorname{Tr}\Pi\,[(\mathcal F\circ\mathcal E_{\rm true})\otimes\mathrm{id}_R](\rho)
 &-\operatorname{Tr}\Pi\,[(\mathcal F\circ\mathcal E_{\rm bdry})\otimes\mathrm{id}_R](\rho)\bigr|\\
 &\le 1-e^{-\Gamma_\perp\tau}.
\end{split}
\label{eq:shutdown-bound-test}
\end{equation}
\end{proposition}
\begin{proof}
Shift time so that $t_0=0$. Write $\theta(t,s)=\int_s^t u(t')\,dt'$ and $U(t,s)=e^{-i\theta(t,s)P/2}$, so that $U(\tau,0)=e^{-i\pi P/2}=-iP$. With $\Gamma=\sum_\mu\gamma_\mu$ and $\mathcal K=\sum_\mu\gamma_\mu\operatorname{Ad}_{E_\mu}$, the generator during the pulse is $u(t)\mathcal H_P+\mathcal K-\Gamma\,\mathrm{id}$. The term $-\Gamma\,\mathrm{id}$ commutes with every superoperator and contributes the factor $e^{-\Gamma\tau}$. Expanding the remaining propagator in $\mathcal K$ in the interaction picture of the pulse gives the Dyson series
\begin{equation}
\begin{split}
 \mathcal E_{\rm true}=e^{-\Gamma\tau}\sum_{n\ge0}\sum_{\mu_1,\dots,\mu_n}
 &\int_{0<t_1<\dots<t_n<\tau}\prod_{i=1}^n\gamma_{\mu_i}\,dt_i\;\operatorname{Ad}_{W},\\
 W={}&U(\tau,t_n)E_{\mu_n}U(t_n,t_{n-1})\cdots E_{\mu_1}U(t_1,0),
\end{split}
\label{eq:dyson-unravelling}
\end{equation}
which converges absolutely because $\|\mathcal K\|_\diamond\le\Gamma$ and the $n$-th term is bounded by $e^{-\Gamma\tau}(\Gamma\tau)^n/n!$. The weights $e^{-\Gamma\tau}\prod_i\gamma_{\mu_i}\,dt_i$ are nonnegative and sum to one: they define a probability measure $\mathbb P$ on jump histories $h=(n;t_1,\mu_1;\dots;t_n,\mu_n)$, the law of independent Poisson processes of rates $\gamma_\mu$. Hence $\mathcal E_{\rm true}=\int\operatorname{Ad}_{W_h}\,d\mathbb P(h)$. The same expansion with $u\equiv0$ gives $e^{\tau\mathcal L_N}=\int\operatorname{Ad}_{E_{\mu_n}\cdots E_{\mu_1}}\,d\mathbb P(h)$ with the same measure, and therefore
\begin{equation}
 \mathcal E_{\rm bdry}=\int\operatorname{Ad}_{PE_{\mu_n}\cdots E_{\mu_1}}\,d\mathbb P(h).
\end{equation}

Let $G$ be the set of histories containing at least one jump $\mu_i$ with $\{E_{\mu_i},P\}=0$. Outside $G$, every $E_{\mu_i}$ commutes with $P$ and hence with every $U(t,s)$, so $W_h=E_{\mu_n}\cdots E_{\mu_1}U(\tau,0)=-iE_{\mu_n}\cdots E_{\mu_1}P$. Global phases drop out of $\operatorname{Ad}$ and $P$ commutes with the jumps, so $\operatorname{Ad}_{W_h}=\operatorname{Ad}_{PE_{\mu_n}\cdots E_{\mu_1}}$: the two integrands coincide outside $G$, and
\begin{equation}
 \mathcal E_{\rm true}-\mathcal E_{\rm bdry}
 =\int_G\bigl(\operatorname{Ad}_{W_h}-\operatorname{Ad}_{PE_{\mu_n}\cdots E_{\mu_1}}\bigr)\,d\mathbb P(h).
\end{equation}
Summing the weights of the histories whose marks all commute with $P$, with $\Gamma_\parallel=\Gamma-\Gamma_\perp$,
\begin{equation}
 \mathbb P(G^{c})=e^{-\Gamma\tau}\sum_{n\ge0}\frac{(\Gamma_\parallel\tau)^n}{n!}=e^{-\Gamma_\perp\tau}.
\end{equation}
Every unitary channel has unit diamond norm, so each integrand has diamond norm at most two. The triangle inequality, applied to the absolutely convergent sum of integrals, gives $\|\mathcal E_{\rm true}-\mathcal E_{\rm bdry}\|_\diamond\le2\,\mathbb P(G)=2(1-e^{-\Gamma_\perp\tau})$, which is Eq.~\eqref{eq:shutdown-bound}.

For Eq.~\eqref{eq:shutdown-bound-test}, let $\Delta=\mathcal E_{\rm true}-\mathcal E_{\rm bdry}$ and $X=[(\mathcal F\circ\Delta)\otimes\mathrm{id}_R](\rho)$. $X$ is Hermitian and traceless, so $X=X_+-X_-$ with $X_\pm\ge0$ and $\operatorname{Tr}X_+=\operatorname{Tr}X_-=\tfrac12\|X\|_1$; with $0\le\Pi\le I$ this gives $|\operatorname{Tr}\Pi X|\le\tfrac12\|X\|_1$. A positive trace-preserving map does not increase the trace norm of a Hermitian operator, so $\|X\|_1\le\|(\Delta\otimes\mathrm{id}_R)(\rho)\|_1\le\|\Delta\|_\diamond$. The last inequality holds for a reference of any dimension, because the diamond norm is already attained with a reference as large as the register.
\end{proof}
The bound is conservative, since isotropic depolarization satisfies the exact identity although $\Gamma_\perp>0$. It does not cover a tempered posterior, feedback from a residual measurement record, or arbitrarily many future adaptive corrections, which we assess by closed-loop simulation.

\paragraph*{Dense checks.}
A validation script propagates full density matrices on register plus reference with arbitrary mixed initial states, Pauli noise, and positive QND maps, alternating detection (efficiency 0.63, strength 40) with variable-amplitude area-$\pi$ pulses of duration 0.008 to 0.010. Six pulses are tested for the three-qubit code and thirty for the five-qubit code, the latter selecting axes from the preceding record. The largest five-qubit syndrome-probability discrepancy is $2.68\times10^{-14}$ during detection and $2.44\times10^{-14}$ at completion, the largest pulse-channel trace-distance residual is $1.02\times10^{-14}$, and the three-qubit residuals are below $4.7\times10^{-15}$. Single-qubit counterexamples separate the assumptions: a covariant case agrees to $8.0\times10^{-16}$ in superoperator Frobenius norm, whereas anisotropic noise, residual monitoring, and a pulse-area error give nonzero residuals.

\section{Simulation and estimator updates}
\label{app:numerics}
Physical trajectories use Pauli unitary rotations, hidden Pauli noise, and positive measurement maps. Monitoring a Hermitian Pauli $S$ uses the normalized map $M\propto\exp(\sqrt{\eta\kappa}\,S\,dY)$ and an unobserved dephasing factor $e^{-2(1-\eta)\kappa\delta t}$ between opposite eigenspaces. Split composition introduces finite-step error but never requires repairing a negative density matrix, and all physical jumps and unobserved measurement events are hidden from the estimator. Kernels are compiled with NumPy/Numba.

For three-qubit bit flips, with error-amplitude index $e\in\{0,\ldots,7\}$ and logical-$X$ partner $e\oplus7$, the minimum-weight recovery gives $F_e^{(\mathrm{traj})}=\sum_{e\in\{0,1,2,4\}}|\psi_e|^2$ and $F^{(\mathrm{traj})}=|\psi_0|^2$. For five qubits, in the 64-dimensional basis $E_aL_\alpha|\Phi_L\rangle$ with sixteen syndrome representatives and $L_\alpha\in\{I,X_L,Y_L,Z_L\}$, $F_e^{(\mathrm{traj})}=\sum_a|\psi_{a,I}|^2$ and $F^{(\mathrm{traj})}=|\psi_{0,I}|^2+|\psi_{0,Z}|^2$; the latter is the joint projector onto $|0_L0_R\rangle$ and $|1_L1_R\rangle$. Both formulas are checked on arbitrary Bell-space vectors, and the moment estimator is checked against a full $32\times32$ density update.

\subsection{Positive split filter}
\label{app:split}
For a measured Pauli $S$, write $m=\langle S\rangle$, $z=2\sqrt{\eta\kappa}\,dY$, and $d=\cosh z+m\sinh z$. The split measurement update of a retained moment $x_O$ is
\begin{equation}
 x_O^+=\begin{cases}
 [\cosh z\,x_O+\sinh z\,x_{SO}]/d,&[S,O]=0,\\
 e^{-2(1-\eta)\kappa\delta t}x_O/d,&\{S,O\}=0,
 \end{cases}
\end{equation}
with Pauli product phases included in $x_{SO}$. The Hamiltonian update uses sine--cosine rotations of retained commutator pairs, and the noise their exponential damping. Omitted Hamiltonian couplings are projected out in the nine-coordinate model; the exact 15- and 495-moment models keep their full generated spaces.

\subsection{Tempered sixteen-class estimator}
For the tempered five-qubit branch, with class probabilities $p_a$ and syndrome signs $s_{ja}=\pm1$, the prediction is $\widetilde p_a=p_a+\delta t\,\gamma_5\left(\sum_b p_b-16p_a\right)$. After flooring at $10^{-15}$ and normalization, the record update multiplies by
\begin{equation}
 w_a=\exp\!\left[2\beta c\sqrt{\eta\kappa}
 \sum_j s_{ja}\,dY_j-\max_b\ell_b\right],
\end{equation}
where $\ell_b$ is the unshifted exponent, $\beta=0.5$, and $c=0.9$, with a $10^{-300}$ floor before final normalization. The initial unnormalized weights are $(1,0.001,\ldots,0.001)$. During shutdown the likelihood adds no information while the hazard prediction continues, and Eq.~\eqref{eq:completion} is applied after the last substep of the completed pulse. This posterior is approximate during a partial coherent rotation.

\subsection{Gradient-law ignition rule}
The Ahn-type law maximizes the instantaneous drift of the conditional codespace population under componentwise bounded drives. With $P_C$ the codespace projector,
\begin{equation}
 g_q=\frac{i}{2}\Tr([P_q,P_C]\widehat\rho),\qquad
 u_q=u_{\max}\operatorname{sgn}(g_q).
\end{equation}
When $|g_q|<10^{-12}$ and the estimated error-class probability exceeds the codespace probability, the code applies a positive drive on channel $q$. The Geremia adaptation uses the same decision with the first-order truncated estimator.

\section{Five-qubit resources and numerical checks}
\label{app:unified-checks}
\begin{table*}[t]
\caption{Resources and integration sensitivity in the five-qubit benchmark. Resources are means from the main test and are not equalized. Fine-step endpoints use an independent ensemble and include sampling changes as well as integration sensitivity.}
\begin{ruledtabular}
\begin{tabular}{lrrrr}
Method & $B_\kappa$ & $A_u$ & Fine-step $F_e(3)$ & Fine-step $F(3)$\\
Shutdown, calibrated & 4610.2 & 46.6 & $0.6875\pm0.0584$ & $0.8750\pm0.0417$\\
Shutdown, tempered & 4608.0 & 47.1 & $0.6250\pm0.0610$ & $0.8594\pm0.0438$\\
Floored LQR, fixed & 4800.0 & 141.6 & $0.3125\pm0.0584$ & $0.5628\pm0.0625$\\
Floored LQR, switched & 4378.3 & 103.3 & $0.4219\pm0.0622$ & $0.5469\pm0.0627$\\
Shifted-design LQR & 4258.0 & 123.4 & $0.3109\pm0.0560$ & $0.5024\pm0.0613$\\
Dynamic inverse & 1200.0 & 169.6 & $0.3710\pm0.0602$ & $0.5196\pm0.0611$\\
Ahn & 1200.0 & 16306.5 & $0.2626\pm0.0336$ & $0.5040\pm0.0364$\\
Geremia first order$^{\dagger}$ & 2400.0 & 4194.8 & $0.2415\pm0.0312$ & $0.3698\pm0.0424$\\
Borah-MBE extension & 4800.0 & 134.8 & $0.2656\pm0.0556$ & $0.5159\pm0.0629$\\
Measurement-on pulse & 4800.0 & 281.7 & $0.1875\pm0.0492$ & $0.3360\pm0.0589$\\
No feedback / codespace LQR & 4800.0 & 0.0 & $0.2344\pm0.0534$ & $0.0312\pm0.0219$\\
\end{tabular}
\end{ruledtabular}
\end{table*}
No trajectory failed in the main test or the refinement. Syndrome-probability violations above $10^{-8}$ occurred only for Geremia first order in the main test (4 trajectories; maximum 0.133). The full auxiliary Pauli-coordinate map agrees with dense Kraus and unitary propagation to $1.2\times10^{-14}$; the native branch agrees after the declared sector projection, which this does not validate as an exact switched filter. The tempered update agrees with the original formula to $2.8\times10^{-17}$. The calibrated update agrees with a dense register--reference calculation to $2.93\times10^{-14}$ during detection and $1.52\times10^{-14}$ at completion over all fifteen axes, which validates the split-map implementation. For calibrated shutdown, the paired 128-trajectory half-step change is $+0.0216$, interval $[-0.0454,0.0886]$, in $F_e(3)$ and $+0.0058$, interval $[-0.0497,0.0614]$, in $F(3)$.

\section{Three-qubit comparison with two monitored generators}
\label{app:three-two-gen}
We repeat Sec.~\ref{sec:three-unified} with only $Z_1Z_2$ and $Z_2Z_3$ monitored, the minimal configuration of Sec.~\ref{sec:methods}. The multiplicities are $w=(1,2,1)$ and the sector matrices are those of Eq.~\eqref{eq:sector-bilinear}; the fixed-channel filter keeps six coordinates and the calibrated filter four classes with a two-record likelihood. Grids, seeds, ensemble sizes, and selection rules are those of Sec.~\ref{sec:three-unified}, and every family is retrained. The published Borah protocol, which monitors three parities, is included from the main comparison with the same seeds; the same local-$Z$ rule on two generator records is a separate arm. Dense register--reference checks validate the filter maps for both configurations to below $4\times10^{-13}$.

\begin{figure*}[t!]
\includegraphics[width=\textwidth]{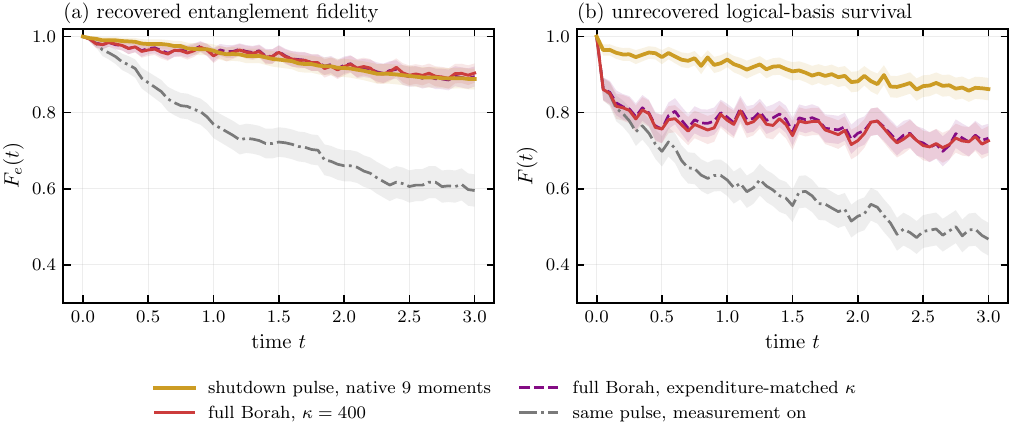}
\caption{Earlier three-qubit full-protocol benchmark. $F_e$ includes the common ideal terminal recovery; $F$ is unrecovered logical-basis survival. 512 held-out trajectories per policy at $\eta=0.75$, $h=10^{-4}$, $\delta t=1.25\times10^{-5}$; shading is pointwise 95\% uncertainty. The nine-coordinate split pulse reproduces the metrics of its full-density counterpart exactly in this ensemble. Measurement expenditure is calibrated only for the labeled budget comparator.}
\label{fig:earl_comp}
\end{figure*}

\begin{table*}[t]
\caption{Frozen settings and held-out three-qubit endpoints at $T=3$ ($N=512$ per method, $\eta=0.75$, $\gamma=1$, monitored generators $Z_1Z_2$ and $Z_2Z_3$; the published Borah protocol additionally monitors $Z_1Z_3$~\cite{borah_measurement_2022}). Columns as in Table~\ref{tab:three-unified}.}
\label{tab:three-unified-2g}
\begin{ruledtabular}
\begin{tabular}{lrrrrllrr}
Method & $\kappa_{\rm det}$ & $\kappa_{\rm cor}$ & $u_{\max}^{\rm used}$ & $\theta$/$\epsilon$ & $F_e(3)$ & $F(3)$ & $B_\kappa$ & $A_u$\\
Shutdown, calibrated & 400 & 0 & 200 & 0.99 & $0.9062\pm0.0129$ & $0.8683\pm0.0146$ & 2289 & 27.5\\
Shutdown, tempered & 400 & 0 & 200 & 0.95 & $0.8926\pm0.0137$ & $0.8458\pm0.0156$ & 2292 & 26.9\\
Floored LQR, fixed & 400 & 400 & 400 & -- & $0.7617\pm0.0188$ & $0.7156\pm0.0199$ & 2400 & 63.5\\
Floored LQR, switched & 400 & 50 & 400 & -- & $0.7688\pm0.0186$ & $0.7397\pm0.0192$ & 2293 & 53.2\\
Shifted-design LQR & 400 & 50 & 400 & -- & $0.7657\pm0.0186$ & $0.7286\pm0.0193$ & 2269 & 57.4\\
Dynamic inverse & 400 & 50 & 400 & -- & $0.6773\pm0.0205$ & $0.6052\pm0.0210$ & 2109 & 125.7\\
Ahn & 200 & 200 & 400 & -- & $0.6414\pm0.0181$ & $0.6114\pm0.0186$ & 1200 & 3139.5\\
Geremia first order & 400 & 400 & 200 & -- & $0.4888\pm0.0173$ & $0.4185\pm0.0177$ & 2400 & 1616.4\\
Borah, published (3 parities) & 400 & 400 & 400 & 1.05 & $0.9199\pm0.0120$ & $0.8426\pm0.0160$ & 3600 & 79.7\\
Borah rule, 2 generators & 200 & 200 & 200 & 1.05 & $0.8164\pm0.0171$ & $0.7079\pm0.0199$ & 1200 & 55.4\\
Measurement-on pulse & 400 & 400 & 200 & 0.95 & $0.5133\pm0.0221$ & $0.3345\pm0.0207$ & 2400 & 204.2\\
No feedback / codespace LQR & 400 & 400 & 400 & -- & $0.5039\pm0.0221$ & $0.1309\pm0.0149$ & 2400 & 0.0\\
\end{tabular}
\end{ruledtabular}
\end{table*}

Calibrated shutdown gives $F_e(3)=0.9062\pm0.0129$ and $F(3)=0.8683\pm0.0146$ (Table~\ref{tab:three-unified-2g}). With the same two currents it exceeds the Borah rule by $0.0898$, eight-contrast adjusted interval $[0.0445,0.1352]$. Against the published three-current protocol the paired difference is $-0.0137$, interval $[-0.0399,0.0126]$, in $F_e(3)$ and $+0.0256$, interval $[-0.0063,0.0576]$, in $F(3)$, both unresolved, with expenditure 2289 against 3600 and drive area 27.5 against 79.7. Its margins over the three LQR variants lie between 0.137 and 0.145, with adjusted lower bounds of at least 0.088.

Shifted-design LQR gives $F_e(3)=0.7657\pm0.0186$ and matches floored LQR at the same schedule (paired difference $-0.0031$, interval $[-0.0337,0.0274]$). Reduced measurement does not improve floored LQR here: the switched-minus-fixed difference is $0.0071$, adjusted interval $[-0.0378,0.0521]$, because with $w=1$ in the outer sectors the partners are damped at only $2\kappa$. Measurement shutdown raises the tempered pulse by $0.3793$, interval $[0.3293,0.4292]$, and the calibrated filter adds $0.0137$, interval $[-0.0076,0.0350]$. The refinement gives $0.9336\pm0.0156$ for calibrated shutdown and $0.9219\pm0.0168$ for the three-current Borah protocol, a paired difference of $0.0117$, interval $[-0.0250,0.0485]$. No trajectory fails and no violation above $10^{-8}$ occurs.

\paragraph*{Cross-check of the Borah implementation.}
The earlier study of Appendix~\ref{app:three-borah} used independently written code for the same protocol. With its frozen policy ($\kappa=400$, $\lambda=200$, $\epsilon=1.05$), step $1.25\times10^{-5}$, and 512 trajectories, the present kernel gives $F_e(3)=0.8768\pm0.0144$ and the earlier implementation $0.9040\pm0.0129$, with independent random streams. The difference, $-0.0272$, is 1.4 standard errors.

\begin{figure*}[t!]
\includegraphics[width=\textwidth]{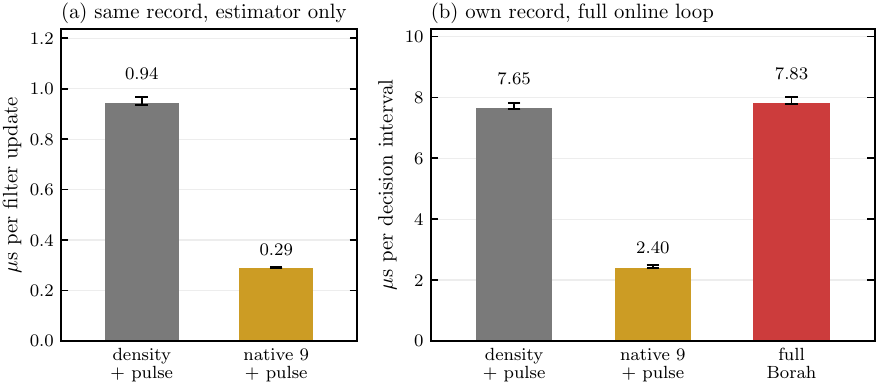}
\caption{Three-qubit online computation. (a) Estimator propagation on identical records. (b) Estimator plus control decision on each protocol's own records. Bars are medians and whiskers interquartile ranges of 30 replay-average times, with compiled double-precision arithmetic and identical integration steps. The $3.26\times$ ratio is hardware and implementation dependent; a second machine gives $4.5\times$ (Sec.~\ref{sec:cost}).}
\label{fig:cost}
\end{figure*}

\section{Earlier three-qubit full-Borah comparison and online cost}
\label{app:three-borah}
This study precedes Sec.~\ref{sec:three-unified}. It monitors three parities and uses the nine-coordinate split filter with a $\theta=0.95$ pulse, and it is the source of the measured online-cost comparison Fig.~\ref{fig:earl_comp}. The model has rate-$\gamma=1$ bit flips, $T=3$, $\eta=0.75$, zero added latency, and the caps $\kappa\le400$, $|u_q|\le400$. The auxiliary density estimator starts in $|000\rangle$ and receives only the currents and applied controls; the published detector drives when $1-\langle Z_q\rangle_{\hat\rho}>\epsilon$, simultaneous drives permitted, with $u_q=2\lambda_q$. An 18-point grid ($\kappa\in\{100,200,400\}$, $u\in\{200,400\}$, $\epsilon\in\{1.05,1.5,1.9\}$, 128 trajectories each) selects $\kappa=400$, $u=200$, $\epsilon=1.05$, followed by a 512-trajectory test at step $1.25\times10^{-5}$ and a 256-trajectory refinement at $6.25\times10^{-6}$, with decision interval $10^{-4}$. A separate six-point search tunes the policy at the pulse's measurement expenditure.

At $T=3$, the nine-coordinate shutdown gives $F_e=0.8887\pm0.0139$ and full Borah $0.9040\pm0.0129$, a paired difference of $-0.0154$, interval $[-0.0428,0.0121]$. The unrecovered fidelity is $0.8619\pm0.0150$ against $0.7259\pm0.0197$, a gain of $0.1360$, interval $[0.0959,0.1761]$. The measurement-on pulse gives $F_e=0.5952\pm0.0217$, so suspending monitoring during recovery adds $0.2935$, interval $[0.2448,0.3423]$. The pulse uses $\E[B_\kappa]=3425.66$ against $3600$ and $\E[A_u]=28.89$ against $128.62$. The expenditure-calibrated Borah policy ($\kappa=380.935$, $B_\kappa=3428.41$) gives $F_e=0.8950\pm0.0135$, an unresolved difference. At the smaller step, shutdown gives $0.8750$ and full Borah $0.8863$, paired interval $[-0.0535,0.0310]$.

\subsection{Measured online computational cost}
\label{sec:cost}
Replaying identical measurement, measurement-strength, and control records isolates estimator propagation: the median cost is $0.289\,\mu$s per update for the nine-coordinate split filter and $0.943\,\mu$s for the full density estimator, a factor of 3.26 (Fig.~\ref{fig:cost}); estimator plus decision on common records gives a factor of about 3.15. Replaying each protocol on its own closed-loop records costs $2.40\,\mu$s per decision interval (eight filter updates) for the pulse and $7.83\,\mu$s for full Borah, again 3.26. The full-density pulse costs $7.65\,\mu$s, so most of the difference comes from estimation, and the exact 15-moment pulse takes $2.57\,\mu$s: most of the saving comes from using a moment representation, and the reduction from fifteen to nine coordinates adds a smaller one.

The benchmark uses Python 3.12.14, NumPy 2.3.5, SciPy 1.17.0, and Numba 0.67.0 on one pinned AMD EPYC 9V74 core, with interleaved randomized jobs, subnormals flushed to zero, and 240000 filter updates and 30000 decision intervals per record. Each median pools 15 repetitions on each of two records after warmup. The density implementation exploits Pauli permutations. Timing excludes plant simulation, random-number generation, compilation, file I/O, acquisition, and actuation. On a second machine (one Intel Xeon Silver 4210 core, Numba 0.62) the ratios are 4.71 and 4.52, with absolute times about four times longer, so the size of the ratio depends on hardware while its direction is reproduced. The native pulse reproduces all saved per-trajectory metrics of the full-density pulse on both test ensembles, with moment residuals below $4.7\times10^{-14}$ on the replay records. The result is an implementation-specific three-qubit speedup, measured with our implementations rather than the original authors' software. A separate five-qubit kernel microbenchmark gave $0.395\,\mu$s per sixteen-class update against $3.298\,\mu$s for the exact 495-moment update (median ratio 8.35 over nine samples), excluding control rotation.

\newpage

\bibliography{cqec}

@article{ahn_quantum_2004,
	title = {Quantum error correction for continuously detected errors with any number of error channels per qubit},
	volume = {70},
	issn = {1050-2947, 1094-1622},
	url = {http://arxiv.org/abs/quant-ph/0402067},
	doi = {10.1103/PhysRevA.70.024302},
	number = {2},
	urldate = {2026-04-10},
	journal = {Physical Review A},
	author = {Ahn, Charlene and Wiseman, H. M. and Jacobs, Kurt},
	month = aug,
	year = {2004},
	note = {arXiv:quant-ph/0402067},
	pages = {024302},
}

@article{ahn_continuous_2002,
	title = {Continuous quantum error correction via quantum feedback control},
	volume = {65},
	url = {https://link.aps.org/doi/10.1103/PhysRevA.65.042301},
	doi = {10.1103/PhysRevA.65.042301},
	number = {4},
	urldate = {2026-04-10},
	journal = {Physical Review A},
	publisher = {American Physical Society},
	author = {Ahn, Charlene and Doherty, Andrew C. and Landahl, Andrew J.},
	month = mar,
	year = {2002},
	pages = {042301},
}

@article{chase_efficient_2008,
	title = {Efficient feedback controllers for continuous-time quantum error correction},
	volume = {77},
	url = {https://link.aps.org/doi/10.1103/PhysRevA.77.032304},
	doi = {10.1103/PhysRevA.77.032304},
	number = {3},
	urldate = {2026-04-10},
	journal = {Physical Review A},
	publisher = {American Physical Society},
	author = {Chase, Bradley A. and Landahl, Andrew J. and Geremia, JM},
	month = mar,
	year = {2008},
	pages = {032304},
}

@article{Grigoletto2025exactmodelreduction,
  doi = {10.22331/q-2025-07-29-1814},
  url = {https://doi.org/10.22331/q-2025-07-29-1814},
  title = {Exact {M}odel {R}eduction for {C}ontinuous-{T}ime {O}pen {Q}uantum {D}ynamics},
  author = {Grigoletto, Tommaso and Tao, Yukuan and Ticozzi, Francesco and Viola, Lorenza},
  journal = {{Quantum}},
  issn = {2521-327X},
  publisher = {{Verein zur F{\"{o}}rderung des Open Access Publizierens in den Quantenwissenschaften}},
  volume = {9},
  pages = {1814},
  month = jul,
  year = {2025}
}

@article{linowski_spectral_2023,
	title = {Spectral stabilizability},
	volume = {107},
	issn = {2469-9926, 2469-9934},
	url = {https://link.aps.org/doi/10.1103/PhysRevA.107.042218},
	doi = {10.1103/PhysRevA.107.042218},
	number = {4},
	urldate = {2026-04-07},
	journal = {Physical Review A},
		author = {Linowski, Tomasz and Rudnicki, Lukasz and Gneiting, Clemens},
	month = apr,
	year = {2023},
	pages = {042218},
}

@article{borah_measurement_2022,
	title = {Measurement based estimator scheme for continuous quantum error correction},
	volume = {4},
	issn = {2643-1564},
	url = {http://arxiv.org/abs/2203.13519},
	doi = {10.1103/PhysRevResearch.4.033207},
	number = {3},
	urldate = {2026-03-30},
	journal = {Physical Review Research},
	author = {Borah, Sangkha and Sarma, Bijita and Kewming, Michael and Quijandria, Fernando and Milburn, Gerard J. and Twamley, Jason},
	month = sep,
	year = {2022},
	note = {arXiv:2203.13519 [quant-ph]},
	pages = {033207},
}

@article{Reuer_2023,
   title={Realizing a deep reinforcement learning agent for real-time quantum feedback},
   volume={14},
   ISSN={2041-1723},
   url={http://dx.doi.org/10.1038/s41467-023-42901-3},
   DOI={10.1038/s41467-023-42901-3},
   number={1},
   journal={Nature Communications},
   publisher={Springer Science and Business Media LLC},
   author={Reuer, Kevin and Landgraf, Jonas and Fösel, Thomas and O’Sullivan, James and Beltrán, Liberto and Akin, Abdulkadir and Norris, Graham J. and Remm, Ants and Kerschbaum, Michael and Besse, Jean-Claude and Marquardt, Florian and Wallraff, Andreas and Eichler, Christopher},
   year={2023},
   month=nov }

@misc{vanhandel2005,
  title         = {Optimal error tracking via quantum coding and continuous syndrome measurement},
  author        = {van Handel, Ramon and Mabuchi, Hideo},
  year          = {2005},
  eprint        = {quant-ph/0511221},
  archivePrefix = {arXiv},
  note          = {arXiv:quant-ph/0511221},
}

@article{GrigolettoPellegriniTicozzi2025,
  title   = {Quantum Model Reduction for Continuous-Time Quantum Filters},
  author  = {Grigoletto, Tommaso and Pellegrini, Cl{\'e}ment and Ticozzi, Francesco},
  journal = {Annales Henri Poincar{\'e}},
  year    = {2025},
  doi     = {10.1007/s00023-025-01622-7},
  note    = {arXiv:2501.13885},
}

@article{ahn_quantum_2003,
  title   = {Quantum error correction for continuously detected errors},
  author  = {Ahn, Charlene and Wiseman, H. M. and Milburn, G. J.},
  journal = {Physical Review A},
  volume  = {67},
  pages   = {052310},
  year    = {2003},
  doi     = {10.1103/PhysRevA.67.052310},
}

@article{sarovar_practical_2004,
  title   = {Practical scheme for error control using feedback},
  author  = {Sarovar, Mohan and Ahn, Charlene and Jacobs, Kurt and Milburn, Gerard J.},
  journal = {Physical Review A},
  volume  = {69},
  pages   = {052324},
  year    = {2004},
  doi     = {10.1103/PhysRevA.69.052324},
}

@article{sarovar_cooling_2005,
  title   = {Continuous quantum error correction by cooling},
  author  = {Sarovar, Mohan and Milburn, G. J.},
  journal = {Physical Review A},
  volume  = {72},
  pages   = {012306},
  year    = {2005},
  doi     = {10.1103/PhysRevA.72.012306},
}

@article{mabuchi_continuous_2009,
  title   = {Continuous quantum error correction as classical hybrid control},
  author  = {Mabuchi, Hideo},
  journal = {New Journal of Physics},
  volume  = {11},
  pages   = {105044},
  year    = {2009},
  doi     = {10.1088/1367-2630/11/10/105044},
}

@article{kerckhoff_designing_2010,
  title   = {Designing Quantum Memories with Embedded Control: Photonic Circuits for Autonomous Quantum Error Correction},
  author  = {Kerckhoff, Joseph and Nurdin, Hendra I. and Pavlichin, Dmitri S. and Mabuchi, Hideo},
  journal = {Physical Review Letters},
  volume  = {105},
  pages   = {040502},
  year    = {2010},
  doi     = {10.1103/PhysRevLett.105.040502},
}

@article{ippoliti_perturbative_2015,
  title   = {Perturbative approach to continuous-time quantum error correction},
  author  = {Ippoliti, Matteo and Mazza, Leonardo and Rizzi, Matteo and Giovannetti, Vittorio},
  journal = {Physical Review A},
  volume  = {91},
  pages   = {042322},
  year    = {2015},
  doi     = {10.1103/PhysRevA.91.042322},
}

@article{hsu_method_2016,
  title   = {Method for quantum-jump continuous-time quantum error correction},
  author  = {Hsu, Kung-Chuan and Brun, Todd A.},
  journal = {Physical Review A},
  volume  = {93},
  pages   = {022321},
  year    = {2016},
  doi     = {10.1103/PhysRevA.93.022321},
}

@article{cardona_continuous_2019,
  title   = {Continuous-time Quantum Error Correction with Noise-assisted Quantum Feedback},
  author  = {Cardona, Gerardo and Sarlette, Alain and Rouchon, Pierre},
  journal = {IFAC-PapersOnLine},
  volume  = {52},
  pages   = {198--203},
  year    = {2019},
}

@article{mohseninia_alwayson_2020,
  title   = {Always-On Quantum Error Tracking with Continuous Parity Measurements},
  author  = {Mohseninia, Razieh and Yang, Jing and Siddiqi, Irfan and Jordan, Andrew N. and Dressel, Justin},
  journal = {Quantum},
  volume  = {4},
  pages   = {358},
  year    = {2020},
  doi     = {10.22331/q-2020-11-04-358},
}

@article{atalaya_baconshor_2020,
  title   = {Error-correcting {Bacon-Shor} code with continuous measurement of noncommuting operators},
  author  = {Atalaya, Juan and Korotkov, Alexander N. and Whaley, K. Birgitta},
  journal = {Physical Review A},
  volume  = {102},
  pages   = {022415},
  year    = {2020},
  doi     = {10.1103/PhysRevA.102.022415},
}

@article{atalaya_timedependent_2021,
  title   = {Continuous quantum error correction for evolution under time-dependent {Hamiltonians}},
  author  = {Atalaya, J. and Zhang, S. and Niu, M. Y. and Babakhani, A. and Chan, H. C. H. and Epstein, J. M. and Whaley, K. B.},
  journal = {Physical Review A},
  volume  = {103},
  pages   = {042406},
  year    = {2021},
  doi     = {10.1103/PhysRevA.103.042406},
}

@article{convy_machine_2022,
  title   = {Machine learning for continuous quantum error correction on superconducting qubits},
  author  = {Convy, Ian and Liao, Haoran and Zhang, Song and Patel, Sahil and Livingston, William P. and Nguyen, Ho Nam and Siddiqi, Irfan and Whaley, K. Birgitta},
  journal = {New Journal of Physics},
  volume  = {24},
  pages   = {063019},
  year    = {2022},
  doi     = {10.1088/1367-2630/ac66f9},
}

@article{convy_logarithmic_2022,
  title   = {A Logarithmic {Bayesian} Approach to Quantum Error Detection},
  author  = {Convy, Ian and Whaley, K. Birgitta},
  journal = {Quantum},
  volume  = {6},
  pages   = {680},
  year    = {2022},
  doi     = {10.22331/q-2022-04-04-680},
}

@misc{lanka_optimizing_2025,
  title         = {Optimizing continuous-time quantum error correction for {Markovian} and non-{Markovian} noise models},
  author        = {Lanka, Anirudh and Hegde, Shashank and Brun, Todd A.},
  year          = {2025},
  eprint        = {2506.21707},
  archivePrefix = {arXiv},
  note          = {arXiv:2506.21707},
}

@misc{halaski_obstacles_2026,
  title         = {Obstacles to Continuous Quantum Error Correction via Parity Measurements},
  author        = {Halaski, Anton and Koch, Christiane P.},
  year          = {2026},
  eprint        = {2603.02106},
  archivePrefix = {arXiv},
  note          = {arXiv:2603.02106},
}

@article{livingston_experimental_2022,
  title   = {Experimental demonstration of continuous quantum error correction},
  author  = {Livingston, William P. and Blok, Machiel S. and Flurin, Emmanuel and Dressel, Justin and Jordan, Andrew N. and Siddiqi, Irfan},
  journal = {Nature Communications},
  volume  = {13},
  pages   = {2307},
  year    = {2022},
  doi     = {10.1038/s41467-022-29906-0},
}

@article{vijay_stabilizing_2012,
  title   = {Stabilizing {Rabi} oscillations in a superconducting qubit using quantum feedback},
  author  = {Vijay, R. and Macklin, C. and Slichter, D. H. and Weber, S. J. and Murch, K. W. and Naik, R. and Korotkov, A. N. and Siddiqi, I.},
  journal = {Nature},
  volume  = {490},
  pages   = {77--80},
  year    = {2012},
  doi     = {10.1038/nature11505},
}

@article{salathe_lowlatency_2018,
  title   = {Low-Latency Digital Signal Processing for Feedback and Feedforward in Quantum Computing and Communication},
  author  = {Salath{\'e}, Y. and Kurpiers, P. and Karg, T. and Lang, C. and Andersen, C. K. and Akin, A. and Krinner, S. and Eichler, C. and Wallraff, A.},
  journal = {Physical Review Applied},
  volume  = {9},
  pages   = {034011},
  year    = {2018},
  doi     = {10.1103/PhysRevApplied.9.034011},
}

@article{mcclure_rapid_2016,
  title   = {Rapid Driven Reset of a Qubit Readout Resonator},
  author  = {McClure, Douglas T. and Paik, Hanhee and Bishop, Lev S. and Steffen, Matthias and Chow, Jerry M. and Gambetta, Jay M.},
  journal = {Physical Review Applied},
  volume  = {5},
  pages   = {011001},
  year    = {2016},
  doi     = {10.1103/PhysRevApplied.5.011001},
}

@book{wiseman_milburn_2009,
  title     = {Quantum Measurement and Control},
  author    = {Wiseman, Howard M. and Milburn, Gerard J.},
  publisher = {Cambridge University Press},
  address   = {Cambridge},
  year      = {2009},
}

@article{jacobs_steck_2006,
  title   = {A straightforward introduction to continuous quantum measurement},
  author  = {Jacobs, Kurt and Steck, Daniel A.},
  journal = {Contemporary Physics},
  volume  = {47},
  number  = {5},
  pages   = {279--303},
  year    = {2006},
  doi     = {10.1080/00107510601101934},
}

@article{bouten_introduction_2007,
  title   = {An Introduction to Quantum Filtering},
  author  = {Bouten, Luc and van Handel, Ramon and James, Matthew R.},
  journal = {SIAM Journal on Control and Optimization},
  volume  = {46},
  number  = {6},
  pages   = {2199--2241},
  year    = {2007},
  doi     = {10.1137/060651239},
}

@article{benoist_exponential_2017,
  title   = {Exponential Stability of Subspaces for Quantum Stochastic Master Equations},
  author  = {Benoist, Tristan and Pellegrini, Cl{\'e}ment and Ticozzi, Francesco},
  journal = {Annales Henri Poincar{\'e}},
  volume  = {18},
  number  = {6},
  pages   = {2045--2074},
  year    = {2017},
  doi     = {10.1007/s00023-017-0556-3},
}

@article{facchi_quantum_2002,
  title   = {Quantum {Zeno} Subspaces},
  author  = {Facchi, P. and Pascazio, S.},
  journal = {Physical Review Letters},
  volume  = {89},
  pages   = {080401},
  year    = {2002},
  doi     = {10.1103/PhysRevLett.89.080401},
}

@phdthesis{gottesman_stabilizer_1997,
  title  = {Stabilizer Codes and Quantum Error Correction},
  author = {Gottesman, Daniel},
  school = {California Institute of Technology},
  year   = {1997},
  note   = {arXiv:quant-ph/9705052},
}

@article{laflamme_perfect_1996,
  title   = {Perfect Quantum Error Correcting Code},
  author  = {Laflamme, Raymond and Miquel, Cesar and Paz, Juan Pablo and Zurek, Wojciech Hubert},
  journal = {Physical Review Letters},
  volume  = {77},
  pages   = {198--201},
  year    = {1996},
  doi     = {10.1103/PhysRevLett.77.198},
}

@article{bennett_mixed_1996,
  title   = {Mixed-state entanglement and quantum error correction},
  author  = {Bennett, Charles H. and DiVincenzo, David P. and Smolin, John A. and Wootters, William K.},
  journal = {Physical Review A},
  volume  = {54},
  pages   = {3824--3851},
  year    = {1996},
  doi     = {10.1103/PhysRevA.54.3824},
}

@article{schumacher_sending_1996,
  title   = {Sending entanglement through noisy quantum channels},
  author  = {Schumacher, Benjamin},
  journal = {Physical Review A},
  volume  = {54},
  pages   = {2614--2628},
  year    = {1996},
  doi     = {10.1103/PhysRevA.54.2614},
}

@book{anderson_optimal_1990,
  title     = {Optimal Control: Linear Quadratic Methods},
  author    = {Anderson, Brian D. O. and Moore, John B.},
  publisher = {Prentice Hall},
  address   = {Englewood Cliffs, NJ},
  year      = {1990},
}

@book{elliott_bilinear_2009,
  title     = {Bilinear Control Systems: Matrices in Action},
  author    = {Elliott, David L.},
  series    = {Applied Mathematical Sciences},
  volume    = {169},
  publisher = {Springer},
  address   = {Dordrecht},
  year      = {2009},
}

@article{altafini_modeling_2012,
  title   = {Modeling and Control of Quantum Systems: An Introduction},
  author  = {Altafini, Claudio and Ticozzi, Francesco},
  journal = {IEEE Transactions on Automatic Control},
  volume  = {57},
  number  = {8},
  pages   = {1898--1917},
  year    = {2012},
  doi     = {10.1109/TAC.2012.2195830},
}

@article{blom_interacting_1988,
  title   = {The interacting multiple model algorithm for systems with {Markovian} switching coefficients},
  author  = {Blom, Henk A. P. and Bar-Shalom, Yaakov},
  journal = {IEEE Transactions on Automatic Control},
  volume  = {33},
  number  = {8},
  pages   = {780--783},
  year    = {1988},
  doi     = {10.1109/9.1299},
}

@article{enns_dynamic_1994,
  title   = {Dynamic inversion: an evolving methodology for flight control design},
  author  = {Enns, Dale and Bugajski, Dan and Hendrick, Russ and Stein, Gunter},
  journal = {International Journal of Control},
  volume  = {59},
  number  = {1},
  pages   = {71--91},
  year    = {1994},
}

@book{khasminskii_stochastic_2012,
  title     = {Stochastic Stability of Differential Equations},
  author    = {Khasminskii, Rafail},
  edition   = {2},
  series    = {Stochastic Modelling and Applied Probability},
  volume    = {66},
  publisher = {Springer},
  address   = {Berlin, Heidelberg},
  year      = {2012},
}

@article{wald_sequential_1945,
  title   = {Sequential Tests of Statistical Hypotheses},
  author  = {Wald, A.},
  journal = {The Annals of Mathematical Statistics},
  volume  = {16},
  number  = {2},
  pages   = {117--186},
  year    = {1945},
  doi     = {10.1214/aoms/1177731118},
}

@article{chen2020PI,
  title={Quantum proportional-integral (PI) control},
  author={Chen, Hui and Li, Hanhan and Motzoi, Felix and Martin, Leigh and Whaley, K Birgitta and Sarovar, Mohan},
  journal={New Journal of Physics},
  volume={22},
  number={11},
  pages={113014},
  year={2020},
  publisher={IOP Publishing}
}

@article{martin2015hybrid,
  title={Deterministic generation of remote entanglement with active quantum feedback},
  author={Martin, Leigh and Motzoi, Felix and Li, Hanhan and Sarovar, Mohan and Whaley, K Birgitta},
  journal={Physical Review A},
  volume={92},
  number={6},
  pages={062321},
  year={2015},
  publisher={APS}
}

@article{greenfield2024stabilizing,
  title={Stabilizing two-qubit entanglement with dynamically decoupled active feedback},
  author={Greenfield, Sacha and Martin, Leigh and Motzoi, Felix and Whaley, K Birgitta and Dressel, Justin and Levenson-Falk, Eli M},
  journal={Physical Review Applied},
  volume={21},
  number={2},
  pages={024022},
  year={2024},
  publisher={APS}
}

\end{document}